\documentclass[letter,11pt]{article}

\pdfoutput=1 

\usepackage{jheppub} 

\usepackage[sort&compress]{natbib} 
\usepackage{amsmath}
\usepackage{graphicx}
\usepackage{dcolumn}
\usepackage{bm}
\usepackage{bbm}
\usepackage{epsfig}
\usepackage{slashed}
\usepackage{amssymb}
\usepackage{color}
\usepackage[font=small]{caption}
\usepackage[font=small]{subcaption}
\usepackage{tabulary}
\usepackage{soul}
\usepackage{booktabs}
\makeatletter

\newcommand{\newc}{\newcommand}
\newc{\renewc}{\renewcommand}

\def\beq{\begin{equation}}
\def\eeq{\end{equation}}
\def\bea{\begin{eqnarray}}
\def\eea{\end{eqnarray}}
\def\bitem{\begin{itemize}}
\def\eitem{\end{itemize}}
\def\ba{\begin{array}}
\def\ea{\end{array}}
\def\bal{\begin{align}}
\def\eal{\end{align}}
\def\bi{\begin{itemize}}
\def\ei{\end{itemize}}
\def\lsim{\mathrel{\rlap{\lower4pt\hbox{\hskip1pt$\sim$}}
    \raise1pt\hbox{$<$}}}         
\def\gsim{\mathrel{\rlap{\lower4pt\hbox{\hskip1pt$\sim$}}
    \raise1pt\hbox{$>$}}}
\newc{\red}{\textcolor{red}}
\newc{\blue}{\textcolor{blue}}

\newc{\ie}{{\it i.e.~}}          \newc{\etal}{{\it et al.~}}
\newc{\eg}{{\it e.g.~}}          \newc{\etc}{{\it etc.~}}
\newc{\cf}{{\it c.f.~}}
\newc{\vs}{{\it vs.~}}
\newc{\os}{\mbox{\hspace{4pt}}}
\newc{\us}{\mbox{\hspace{12pt}}}

\renewc{\bar}{\overline}

\newc{\gev}{\,{\rm GeV}}
\newc{\mev}{\,{\rm MeV}}
\newc{\ev}{\,{\rm eV}}
\newc{\kev}{\,{\rm keV}}
\newc{\tev}{\,{\rm TeV}}

\newc{\LM}{\mathcal{L}}
\newc{\SM}{\mathcal{S}}

\newc{\HM}{\mathcal{H}}
\newc{\GM}{\mathcal{G}}
\newc{\OM}{\mathcal{O}}
\newc{\FM}{\mathcal{F}}
\newc{\AM}{\mathcal{A}}
\newc{\BM}{\mathcal{B}}
\newc{\NM}{\mathcal{N}}
\newc{\WM}{\mathcal{W}}
\newc{\ZM}{\mathcal{Z}}

\newc{\Chi}{\mathcal{X}}

\definecolor{red1}{cmyk}{0,1,1,0.1}
\definecolor{blue1}{cmyk}{1,0,0,0}

\title{Timing information at HL-LHC:\\
Complete determination of masses of Dark Matter and Long lived particle}

\author[a]{Zachary Flowers,}
\affiliation[a]{Department of Physics and Astronomy, University of Kansas, Lawrence, KS 66045, USA}

\author[b]{Dong Woo Kang,}
\affiliation[b]{Korea Institute for Advanced Study, Seoul 02455, Korea}
\emailAdd{dongwoodkang@kias.re.kr}

\author[a]{Quinn Meier,}

\author[c]{Seong Chan Park,}
\affiliation[c]{Department of Physics and IPAP, Yonsei University, Seoul 03722,  Korea}
\emailAdd{sc.park@yonsei.ac.kr}

\author[a]{Christopher Rogan}
\emailAdd{crogan@ku.edu}

\preprint{KIAS-P19051}

\abstract{
A long-standing kinematic challenge in data analysis at hadron colliders is the determination of the masses of invisible particles. This issue is particularly relevant in searches for evidence of dark matter production, which remains one of the prominent targets of future collider experiments. In this paper, we show that the additional information from the precision timing measurements, provided by planned detector upgrades during the high-luminosity run of the LHC (HL-LHC), allows for previously unrealizable measurements of invisible particle kinematics. As a concrete example, we focus on the signal of pair produced long-lived particles ($LLP_{1,2}$), each decaying with a displaced vertex to visible ($V_{1,2}$) and invisible ($I_{1,2}$) final state particles, $pp \to LLP_1+LLP_2 \to (V_1+I_1)+(V_2+I_2)$. We explicitly show that the complete kinematics of the invisible particles in such events can be determined with the addition of timing information, and evaluate the precision with which the masses of new long-lived and invisible particles can be determined. }


\begin{document}

\maketitle


\section{Introduction}
After Run-3 of the LHC a significant upgrade is planned for the High Luminosity LHC (HL-LHC)~\cite{HL-LHC} period of operation, where expected instantaneous luminosities are a factor of five times larger than the current LHC nominal conditions. While these more intense conditions will enable the LHC experiments to accumulate more than an order of magnitude larger dataset than all of the preceding LHC runs, they also exacerbate the experimental challenge of the multiple pile-up interactions appearing in each event. To mitigate degradation of LHC event reconstruction performance resulting from the potentially hundreds of pile-up interactions expected in an HL-LHC event, both CMS~\cite{MTD-TDR} and ATLAS~\cite{HGTD-ATLAS} are developing precision timing detectors, significantly expanding the capabilities of the experiments. With detector elements capable of time-stamping charged particles with a precision of order $\sigma_{\rm t} \simeq 30$ ps, these precision timing detectors can not only be used to disambiguate interaction vertices but also introduce the possibility for new approaches to searching for evidence of new physics involving long-lived particles (LLPs) through time-of-flight (ToF) measurements~\cite{Liu:2018wte,Cerri:2018rkm}.

In this paper we further explore applications of timing information in collider event reconstruction, focusing on final states with massive invisible particles (i.e. dark matter) following from the decays of neutral LLPs. Specifically, we explicitly demonstrate how timing information allows for the determination of the masses of invisible particles and neutral LLPs in events where they are singly or pair-produced. In these cases, each LLP decays into visible and invisible particles, with: $pp \to LLP_a+LLP_b \to (V_a+I_a) +(V_b+I_b)$. This signature appears in many well-motivated beyond the standard model (BSM) theories, such as the second lightest supersymmetric particle decaying to the lightest supersymmetric particle (LSP) and the second lightest Kaluza-Klein particle decaying to the lightest Kaluza-Klein particle (LKP), respectively~\cite{Martin:1997ns, Appelquist:2000nn, Kong:2010xk, Park:2009cs, Datta:2005zs,Harland-Lang:2018hmi}. When an LLP is charged, or decays exclusively into visible particles, the kinematics of the LLP can be measured directly. However, when invisible particles are present in the decays of neutral LLPs the kinematics of these particles are severely under-constrained with the measurement capabilities of the existing LHC detectors, as has been studied extensively (see e.g.~\cite{Kawagoe:2003jv, Park:2011vw, Park:2009cs, Meade:2010ji, Cottin:2018hyf,jigsaw}). Here, we demonstrate how the anticipated addition of precision timing measurements can be used to determine the kinematic properties of these previously intractable events in their entirety. 

This paper is structured as follows: 
In section \ref{sec:TOF}, we develop two reconstruction methods for the events with LLPs and their decays to dark matter particles and visible particles using the timing and vertex measurement capabilities of the HL-LHC detectors. The performance of these reconstruction methods is described in section \ref{sec:MC} using simulated decays of LLPs and an emulation of an HL-LHC detector and pile-up conditions in benchmark new physics scenarios. Finally, we conclude in section \ref{sec:conclusion}.

\section{LLP reconstruction using timing information}
\label{sec:TOF}

\begin{figure}[!tbh]
    \centering 
    \includegraphics[width=0.5\textwidth]{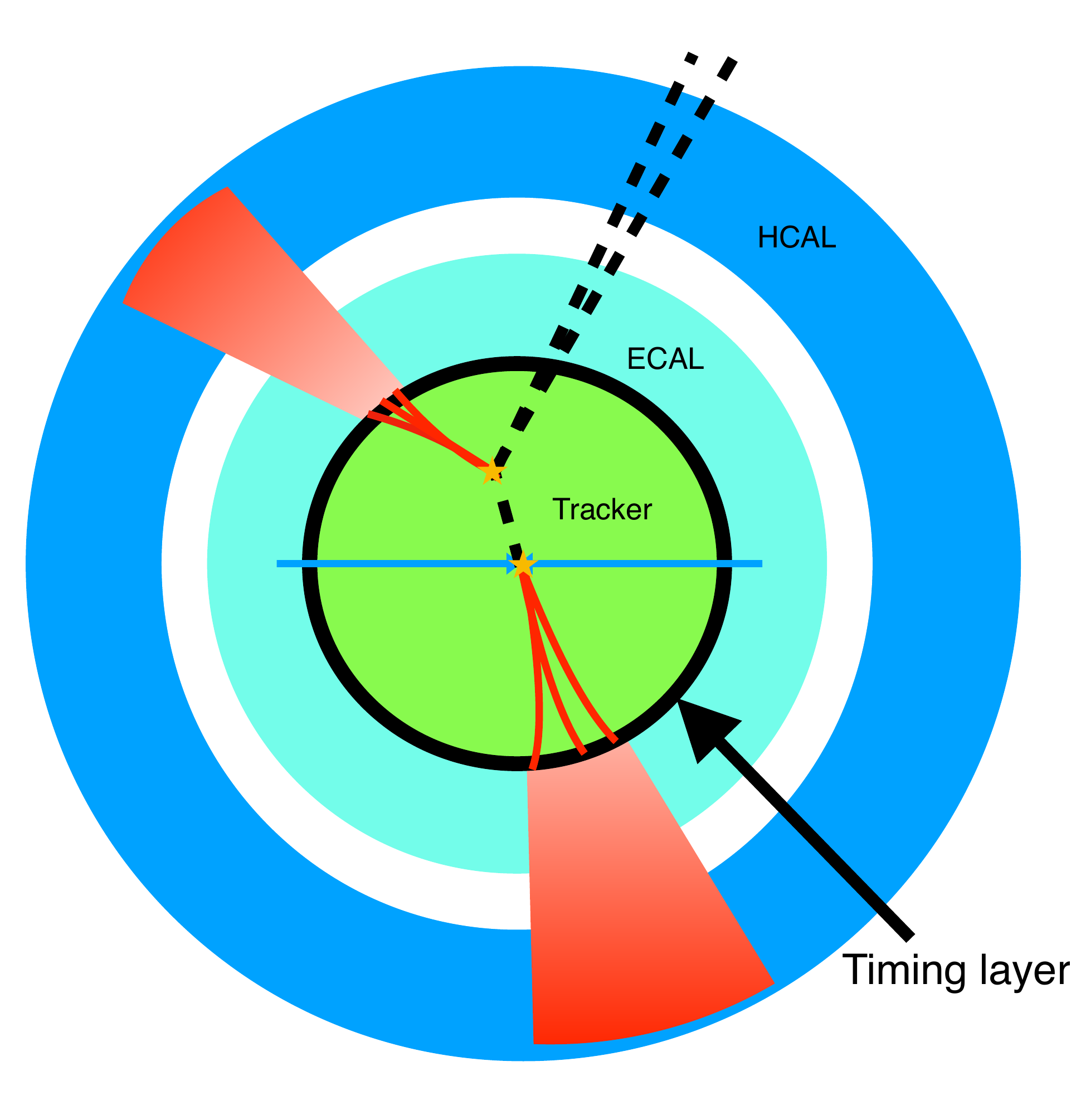}
    \caption{Schematic layout of an HL-LHC detector with the inclusion of a hermetic timing layer.}
    \label{fig:timing_detector}
\end{figure}

We begin by reviewing the general properties of the proposed precision timing detectors at the HL-LHC, focusing in particular on the hermetic design of the CMS timing layer~\cite{MTD-TDR}. An HL-LHC detector with such a timing layer is shown schematically in Fig.~\ref{fig:timing_detector}. The picosecond timing detector will be installed between the inner tracker (green) and the electromagnetic calorimeter (cyan), providing uniform coverage of the barrel as well as the end-cap of the detector. In the case of the ATLAS timing detector, only the end-cap region with be instrumented with precision timing elements~\cite{HGTD-ATLAS}. These timing detectors are capable of detecting minimum ionizing particles (MIPs) with excellent efficiency (nearly 100\%) and time resolution of order 30 ps throughout the lifetime of the HL-LHC. 

At the HL-LHC, during each proton bunch crossing, pile-up interactions are distributed in space and time over the luminous region with dimensions of order 5 cm and 200 ps, respectively. The combination of the inner tracking detector and precision timing layer is able to measure the trajectory of charged particles from initial state radiation (ISR) and the underlying event with an associated time-of-arrival at the timing detector, for each separate interaction. This information can then be used to reconstruct, in space and time, each of the interaction vertices in the event, with the additional time dimension allowing for the disambiguation in the dense HL-LHC environment. 

A neutral LLP produced in an HL-LHC event leaves no corresponding track in the detector, but if it decays to at least two visible, reconstructable particles then its secondary decay vertex can be measured in four dimensions in the same fashion. By associating such a secondary vertex with the corresponding primary interaction vertex in an event then allows for the measurement of the distance and time-of-flight of the LLP and its velocity~\cite{Liu:2018wte}. Here, we examine how this technique can be expanded to incorporate additional kinematic information, like the measured missing transverse energy, to fully reconstruct the decays of these neutral LLPs in events where they are singly or pair-produced.

 We consider the scenario where an LLP travels between 0.1 cm and 1 m and decays in the tracker into visible ($V$) and invisible ($I$) particles. While $V$ and $I$ may correspond to one or more particles in general we consider $V$ (and $I$) as single objects. We further assume that $V$ provides enough measurements to reconstruct the location and time of the secondary vertex where it was produced. Associating this reconstructed secondary vertex with the primary vertex of the interaction the displacement ($\Delta \boldsymbol{r}_{LLP}$), time-of-flight ($\Delta t_{\rm ToF}$), and velocity of the LLP can be calculated:
\begin{eqnarray}
\boldsymbol{\beta}_{LLP} = \frac{\Delta \boldsymbol{r}_{LLP}}{\Delta t_{\rm ToF}}.
\end{eqnarray}
The anticipated uncertainties in the reconstruction of these quantities  ($|\delta (\Delta \boldsymbol{r}_{P})| \lsim (10-30){\rm \mu m}$ and $|\delta(\Delta t_{\rm ToF})| \lsim (30-300) {\rm ps}$~\cite{MTD-TDR}) imply that such a detector would be able to resolve the displaced decays of LLPs over a range of lifetimes and kinematic phase-space, irrespective of the presence of invisible particles (such as dark matter) in these decays.

\subsection{LLP decays to visible and invisible particles ($LLP \to V + I$)}

We first examine the two body decay of a single LLP ($LLP$) to a visible ($V$) and an invisible ($I$) particle ($LLP \rightarrow V + I)$.  The visible particle ($V$), by definition here, is identified unambiguously with its reconstructed 4-momentum in the lab frame: 
 $P_V^{\rm~lab}=(E_{V}^{\rm~lab}, \boldsymbol{p}_{V}^{\rm~lab})$. The mass of $V$ is easily calculable by $m_V^2 = {E_{V}^{\rm~lab}}^2-{\boldsymbol{p}_{V}^{\rm ~lab}}^2$. 
 The 3-velocity of the LLP in the lab frame $\boldsymbol{\beta}_{LLP}^{\rm~lab}$ is also measurable, as explained in the previous section. As the energy of the LLP is not directly measured, it is treated as an unknown kinematic parameter to be determined. Even in the absence of knowledge of the LLP energy, the measured velocity can still be used to evaluate measured 4-vectors in the rest frame of the LLP by boosting from the lab frame by  $\boldsymbol{\beta}_{LLP}^{\rm~lab}$:
\begin{eqnarray}
	E_{V}^{~LLP} = \gamma_{P}^{\rm~lab}\left(E_{V}^{\rm~lab} - \boldsymbol{p}_{V}^{\rm~lab}\cdot \boldsymbol{\beta}_{LLP}^{\rm~lab} \right),
\label{eq:EVLLPa}
\end{eqnarray}
where $\gamma_P^{\rm~lab} = 1/\sqrt{1-\left({\boldsymbol{\beta}^{\rm~lab}_{P}}\right)^2}$ corresponds to the the relativistic gamma factor. Energy-momentum conservation in this reference frame ($p_{LLP}^\mu = p_{V}^\mu + p_{I}^\mu =(m_P,0)$) constrains the 3-momentum of the invisible particle from that of the visible one, with $\boldsymbol{p}_I^{~LLP}=-\boldsymbol{p}_V^{~LLP}$. Furthermore, the energy of the visible particle evaluated in the LLP rest fame is sensitive to the masses of the particles appearing in this decay process, with
\begin{align}
	E_{V}^{~LLP} = \frac{m_{LLP}^{2} - m_{I}^{2} + m_{V}^{2}}{2 m_{LLP}},
\label{eq:EVLLPb}
\end{align}
where $m_{LLP}, m_I$ and  $m_V$ are the masses of the LLP, invisible, and visible particle, respectively. 
Combining Eq.~\ref{eq:EVLLPa} and Eq.~\ref{eq:EVLLPb} yields the expression for the mass of the LLP, 
\begin{align}
	m_{LLP} = E_{V}^{~LLP} + \sqrt{(E_{V}^{~LLP})^{2} + m_{I}^{2} - m_{V}^{2}}.
	\label{eq:mass_LLP}
\end{align}
As $E_V^{~LLP}$ and $m_V$ are measurable, the mass of the invisible particle is the only unknown quantity appearing in this expression.\footnote{Even though the relations of the masses corresponds to the solution of a quadratic equation there's no sign ambiguity involved in Eq.~\ref{eq:mass_LLP}; the sign in front of the square root is chosen by considering the massless limit, $m_{V}\to 0$ and $m_I\to 0$.}

The transverse components of the invisible particle momentum in the lab frame, $\boldsymbol{p}_{I,T}^{\rm~lab}$, can be inferred from the measured missing transverse energy. By equating the sum of the transverse momenta of visible and invisible particles with that of the LLP, the energy of the LLP in the lab frame can be calculated from measured quantities as
\begin{align}\label{eq:pt_conservation}
	\boldsymbol{p}_{LLP, T}^{\rm~lab} &= \boldsymbol{p}_{I,T}^{\rm~lab} + \boldsymbol{p}_{V,T}^{\rm~lab}\nonumber \\
	&= E_{LLP}^{\rm~lab} \boldsymbol{\beta}_{LLP,T}^{\rm~lab} \\
	\Rightarrow E_{LLP}^{\rm~lab} &= \frac{{\boldsymbol{\beta}}_{LLP, T}^{\rm~lab}\cdot \left(\boldsymbol{p}_{I,T}^{\rm~lab} + \boldsymbol{p}_{V,T}^{\rm~lab}\right) }{|\boldsymbol{\beta}_{LLP,T}^{\rm~lab}|^2}.
\end{align}
The mass of the LLP can then be measured according to
 \begin{align}
	m_{LLP} &= \left(\gamma_{LLP}^{\rm~lab} \right)^{-1} E_{LLP}^{\rm ~lab} \label{eq:mass_LLP_gamma} \\
	&= \frac{\sqrt{1-\left({\boldsymbol{\beta}^{\rm ~lab}_{LLP}}\right)^2}}{{|\boldsymbol{\beta}_{LLP,T}^{\rm lab}|^2}} {\boldsymbol{\beta}}_{LLP, T}^{\rm~lab}\cdot \left(\boldsymbol{p}_{I,T}^{\rm~lab} + \boldsymbol{p}_{V,T}^{\rm~lab}\right). 
		\label{eq:mass_LLP_MET}
\end{align}

Similarly, the relation in Eq.~\ref{eq:mass_LLP} can be used to express the mass of the invisible particle:\begin{align}
	m_{I} = \sqrt{m_{LLP}^{2} - 2m_{LLP}\,E_{V}^{~LLP} + m_{V}^{2}}.
	\label{eq:mass_invisible}
\end{align}
While highlighting the measured masses, the above expression corresponds to the complete determination of the masses and momentum of all the particles appearing in the decay process $LLP \to V + I$, facilitated by precision timing measurements.

\subsection{Pair-production of non-identical LLPs ($LLP_a \neq LLP_b$)}
\label{sec:ab}

The approach developed in the previous section can be extended to the pair-production of, potentially non-identical LLPs, denoted as $LLP_{a}$ and $LLP_{b}$.  The energies and momenta of the LLPs are represented by $E_{a}, E_{b}$ and $\boldsymbol{p}_{a}$, $\boldsymbol{p}_{b}$ respectively. 

 For pair-produced LLPs and their decay products, Eq.~\ref{eq:pt_conservation} is now generalized as
\begin{align}
	\boldsymbol{p}_{a, T} + \boldsymbol{p}_{b, T} &= \boldsymbol{p}_{I,T} + \boldsymbol{p}_{V_{a},T} +  \boldsymbol{p}_{V_{b},T} \nonumber \\
	\Rightarrow  E_{a} \boldsymbol{\beta}_{a,T} +  E_{b} \boldsymbol{\beta}_{b,T} &= \boldsymbol{p}_{I,T} + \boldsymbol{p}_{V_{a},T} +  \boldsymbol{p}_{V_{b},T},
	\label{eq:llp2}
\end{align}
where the total transverse momentum of the invisible particles is given by $\boldsymbol{p}_{I,T} = \boldsymbol{p}_{I_{a},T} + \boldsymbol{p}_{I_{b},T}$.  Using the two independent relations from the transverse vector constraint Eq.~\ref{eq:llp2},  the  two unknown energies ($E_{a}$ and $E_{b}$) can be analytically calculated as: 
\begin{align}
E_{a} = \left[\frac{\boldsymbol{\beta}_{b}\times(\boldsymbol{p}_{T}^{miss} + \boldsymbol{p}_{V_{a}} + \boldsymbol{p}_{V_{b}} )\cdot \hat{\boldsymbol{k}}}{\boldsymbol{\beta}_{b} \times \boldsymbol{\beta}_{a} \cdot \hat{\boldsymbol{k}}} \right], \quad E_{b} = \left[\frac{\boldsymbol{\beta}_{a}\times(\boldsymbol{p}_{T}^{miss} + \boldsymbol{p}_{V_{a}} + \boldsymbol{p}_{V_{b}} )\cdot \hat{\boldsymbol{k}}}{\boldsymbol{\beta}_{a} \times \boldsymbol{\beta}_{b} \cdot \hat{\boldsymbol{k}}} \right],
\label{eq:energy_LLP}
\end{align}
where $\hat{\boldsymbol{k}}$ is a unit vector pointing along the beam-line. A detailed derivation of this expression is contained in the Appendix.
With $E_a$ and $E_b$ calculated, the complete 4-momenta of the long-lived and invisible particles are given by:
\begin{align}
p_{a} = (E_{a}, E_{a}\,\boldsymbol{\beta_{a}}),\quad p_{b} = (E_{b}, E_{b}\,\boldsymbol{\beta_{b}}), \quad p_{I_{a}} = p_{a} - p_{V_{a}},\quad p_{I_{b}} = p_{b} - p_{V_{b}},
\end{align}
where $\boldsymbol{\beta}_{a,b}$ is measured using timing information. We emphasize that this derivation is completely generic, in that it can be applied to any system with the same event topology. As these four-vectors are fully-determined, the masses of LLPs and invisible (dark matter) particles can also be calculated; this is one of our main results in this paper.

One may wonder why the addition of timing information is sufficient to determine these quantities of interest. The reason is fairly simple: counting the number of kinematic degrees of freedom (d.o.f.) in the system gives $16$ unknowns corresponding to the four 4-vectors of the two LLPs and two invisible particles. Without timing information, the number of measurable quantities and the conservation conditions are $4+4+2+2+2=14$ where the first two 4's are from the two 4-momenta of visible particles, the next two 2's are from the direction of the displaced vertices and the last 2 are from the total transverse momentum of the invisible particles. The timing information of the two ToF's provide the two additional constraints (in total 16 conditions) to fully-determine the kinematics of the whole system.

\subsection{Production of two identical LLPs ($LLP_a=LLP_b$)}
\label{subsec:identical}


When the pair-produced LLPs are identical, and also with identical decay products, the symmetry constraints on the system reduces the effective number of kinematic unknowns, such that the kinematics of the system can be completely determined even in the absence of timing information.

This can be observed by first considering the relations following from 4-momentum conservation in each branch of the decay processes ($LLP_{i} \to V_{i}+I_{i}$) for $i=a$ or $b$, respectively):
\begin{align}
	p_{I_{a}} &= p_{a} - p_{V_{a}} \nonumber \\
	\Rightarrow m_{I_{a}}^{2} &= m_{a}^{2} + m_{V_{a}}^{2} - \left( 2E_{V_{a}} \sqrt{m_{a}^{2} + |\boldsymbol{p}_{a}|^{2}} - 2\boldsymbol{p}_{V_{a}}\cdot \boldsymbol{p}_{a} \right), \label{eq:curve1}\\
	p_{I_{b}} &= p_{b} - p_{V_{b}} \nonumber  \\
	\Rightarrow m_{I_{b}}^{2} &= m_{b}^{2} + m_{V_{b}}^{2} - \left( 2E_{V_{b}} \sqrt{m_{b}^{2} + |\boldsymbol{p}_{b}|^{2}} - 2\boldsymbol{p}_{V_{b}}\cdot \boldsymbol{p}_{b} \right). \label{eq:curve2}
\end{align}
Enforcing $m_{a} = m_{b} = m_{LLP}$ and $m_{I_{a}} = m_{I_{b}} = m_{I}$ according to assumptions of decay symmetry, the two equations can be combined to yield a quadratic equation for $\Delta \equiv m_{LLP}^2-m_I^2 (>0)$:
\begin{align}
A_{a} \Delta^{2} + 2 B_{a} \Delta + C_{a} = m_{LLP}^{2} = A_{b} \Delta^{2} + 2 B_{b} \Delta+ C_{b}, \label{eq:1st} \\
\therefore (A_a-A_b)\Delta^2 +2 (B_a-B_b)\Delta+ (C_a-C_b)=0, \label{eq:2nd}
\end{align}
where the coefficients for $i=a,b$ are all measurable and can be explicitly written as
\begin{align}
	A_{i} = \frac{1}{4E_{V_{i}}^{2}}, \qquad B_{i} = A_{i}( m_{V_{i}}^{2} + 2\boldsymbol{p}_{V_{i}}\cdot \boldsymbol{p}_{i} ), \qquad C_{i}= \frac{B_{i}^{2}}{A_{i}} -  |\boldsymbol{p}_{i}|^{2}.
\end{align}

For each event, all coefficients ($A$'s, $B$'s and $C$'s) can be calculated from measured quantities, and $\Delta$ can be correspondingly determined. The physical solution is chosen to satisfy the conditions 
$m_{LLP} > m_V +m_I$ and $m_P>0$ and $m_I>0$. Knowledge of $\Delta$, combined with Eq.~
\ref{eq:1st}, is sufficient for determining the LLP and invisible particle masses, which is also one of the results of this paper.

\subsection{Summary of reconstruction methods}

The two equations Eq.~\ref{eq:curve1} and Eq.~\ref{eq:curve2} apply generally to the pair-production of LLPs, irrespective of the availability of timing information, and define two independent algebraic curves, ${\mathcal C}_a$ and ${\mathcal C}_b$, on the two dimensional plane with the coordinates $(m_{LLP_{i}} ,m_{I_{i}})$ for $i=a$ or $i=b$. The point giving the true values of the masses, $(m_{LLP_{i}}, m_{I_{i}})$, will lie on the corresponding curve, ${\mathcal C}_i$, but additional information is required to determine it. 

Two completely solvable scenarios have been described:
\begin{itemize}
\item Pair-production of identical LLPs ($LLP_{a} = LLP_{b}$ and $I_{b} = I_{a}$)
\item Pair-production of non-identical LLPs, using timing ($LLP_{a} \neq LLP_{b}$  and  $I_{a} \neq I_{b}$)
\end{itemize}
For the special case with $m_{LLP_{a}} = m_{LLP_{b}}$ and $m_{I_{a}} = m_{I_{b}}$ the point of intersection of two curves corresponds to the true solution, and can be calculated without timing information. We denote this displaced vertex based reconstruction without timing information simply ``w/o timing reconstruction". A method to reconstruct the 3-momenta of LLP's and invisible particles without timing information has also been developed, with details provided in the Appendix. When timing information is available, assumptions about similarities between the LLPs and invisible particles appearing in the two decays can be relaxed, with the associated measurement of event kinematics denoted ``timing reconstruction". Each of the relevant scenarios are summarized in Table 1.

\vspace{1.0cm}

\begin{table}[h]
\centering
\begin{tabular}{cccccccccc}
\toprule
&  & $m_{LLP_{a}}$ & $m_{LLP_{b}}$ & $m_{I_{a}}$ & $m_{I_{b}}$ & $\boldsymbol{p}_{LLP_{a}}$ & $\boldsymbol{p}_{LLP_{b}}$ & $\boldsymbol{p}_{I_{a}}$ & $\boldsymbol{p}_{I_{b}}$ \\ 
\midrule
Identical LLPs &  w/o timing& $\triangle$ & $\triangle$ & $\triangle$ & $\triangle$ & $\bigcirc$ & $\bigcirc$ & $\bigcirc$ & $\bigcirc$ \\ 
 & timing & $\bigcirc$ & $\bigcirc$ & $\bigcirc$ & $\bigcirc$ & $\bigcirc$ & $\bigcirc$ & $\bigcirc$ & $\bigcirc$ \\ 
\hline 
Non-identical LLPs  & w/o timing & $\times$ & $\times$ & $\times$ & $\times$ & $\bigcirc$ & $\bigcirc$ & $\bigcirc$ & $\bigcirc$ \\ 
 & timing & $\bigcirc$ & $\bigcirc$ & $\bigcirc$ & $\bigcirc$ & $\bigcirc$ & $\bigcirc$ & $\bigcirc$ & $\bigcirc$ \\ 
\bottomrule
\end{tabular} 
\caption{Summary of reconstruction scenarios. The mark $\bigcirc$ ($\times$) indicates whether the system can (cannot) be reconstructed. 
The triangle ($\triangle$) indicates that the system can be reconstruct only with ambiguities.
\label{table:RecoSummary}}
\end{table}

\section{Reconstruction performance in simulated events}
\label{sec:MC}

\begin{figure}[t]
    \centering 
    \includegraphics[width=0.5\textwidth]{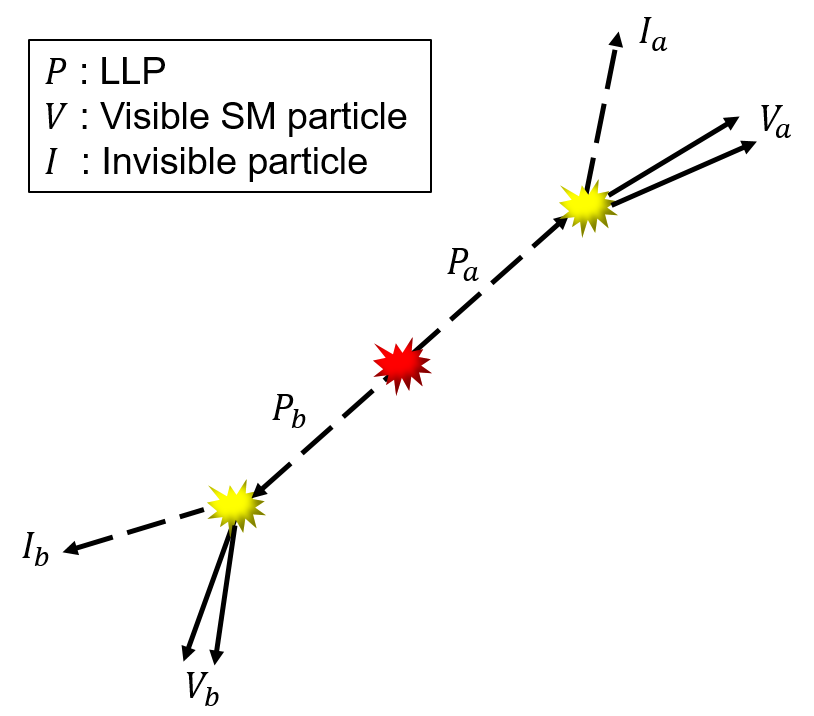}
    \caption{Typical event topology for the LLP pair-production with decays to visible and invisible particles}
    \label{fig:event_toplogy}
\end{figure}

We present two case studies of these LLP reconstruction techniques in scenarios with the pair-production of neutral LLPs independently decaying to visible and invisible particles. The corresponding event topology is illustrated in Fig.~\ref{fig:event_toplogy}. Such a scenario appears in many BSM models, such as gluino decay in GMSB SUSY, slepton decay, gluball decay in hidden sector model and many others~\cite{Chang:2009sv,Graham:2012th,Csaki:2015uza,Allanach:2016pam,Nagata:2017gci,Ito:2017dpm}. In general, requirements on the LLP displacements being experimentally significant in both space and time will remove nearly all SM background contributions to this final state. The characterization and evaluation of remaining background sources is left for future studies, with this paper focusing exclusively on signal events.

The first case study compares the performance of the ``w/o timing reconstruction" and ``timing reconstruction" methods in a generic scenario of LLP pair production and decays, with and without identical LLPs. We further explore the ``timing reconstruction" approach in a second case study, evaluating the precision of measured masses as a function of LLP lifetimes, particle masses, and timing detector resolution.

\subsection{Comparison of reconstruction methods}

To compare the ``w/o timing reconstruction" and ``timing reconstruction" reconstruction methods, simulated signal events are evaluated using a toy representation of an HL-LHC detector. Here, generator-level 4-vectors are smeared according to a simple model of a detector, with the momentum resolution of visible particles taken to be 2\%. The experimental resolutions of LLP displacements in space and time are assumed to have resolutions of 12 $\mu$m and 30 ps, respectively. Events are generated using \texttt{MG5aMC}~\cite{Alwall:2014hca}, with particle decays are simulated with \texttt{Pythia8}~\cite{Sjostrand:2006za, Sjostrand:2007gs,Sjostrand:2014zea}.

\subsubsection{$LLP_{a} =LLP_{b}$  and  $I_{a} = I_{b}$}

We first consider a scenario with identical LLPs and decays, choosing $m_{\rm LLP} = 400$ GeV and $m_{\rm Inv} = 200$ GeV, with a LLP lifetime of $c\tau \approx \mathcal{O}(100) $mm. Using both reconstruction methods, the masses $M_{\rm LLP}$ and $M_{\rm Inv}$ can be evaluated.  Of note in this analysis is the absence of significant combinatoric ambiguities as the two {\it distinguishable} displaced vertices are independently identified and measured. 

\begin{figure}[t]
    \centering 
  \includegraphics[width=0.32\textwidth]{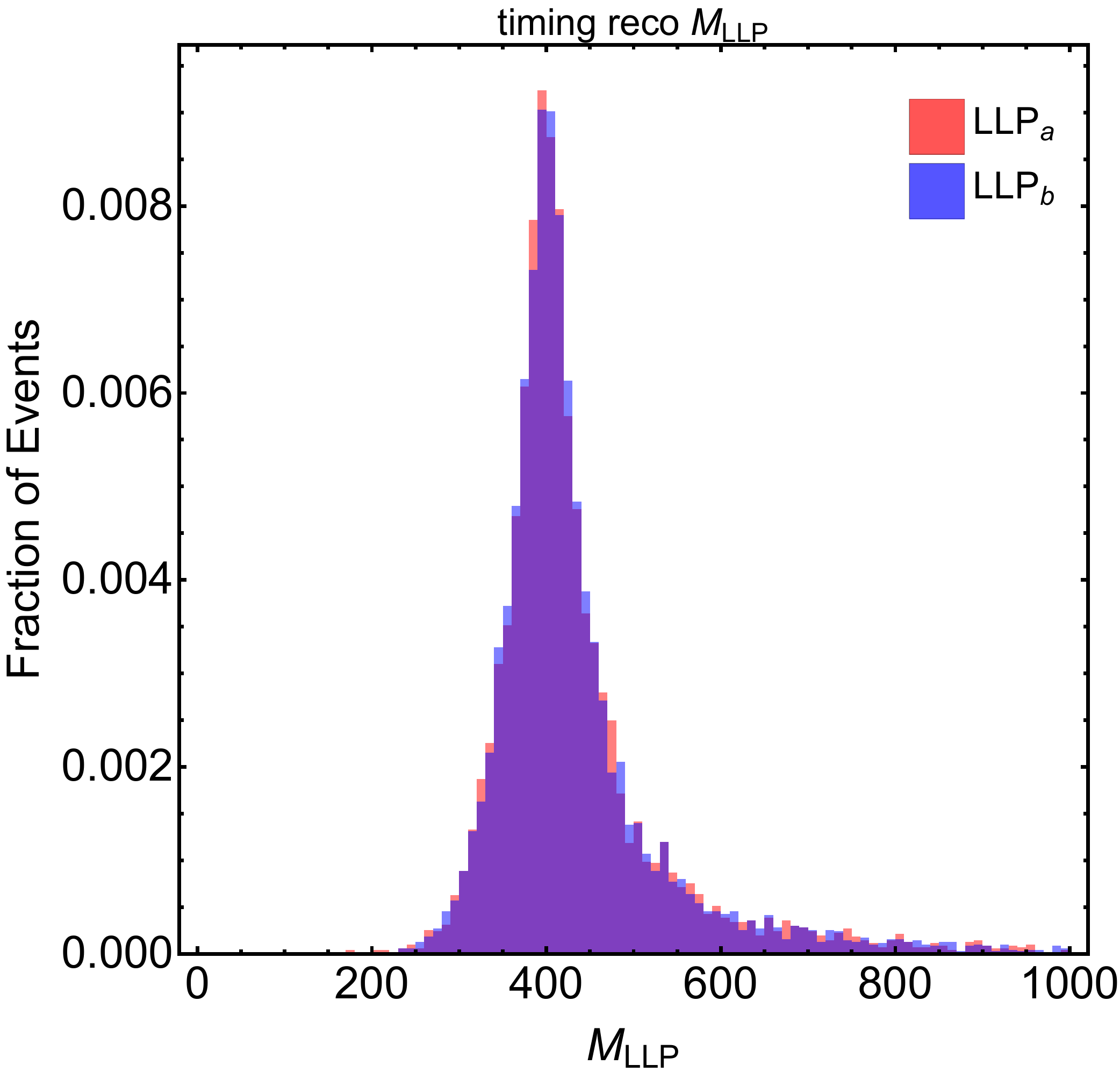}
    \includegraphics[width=0.32\textwidth]{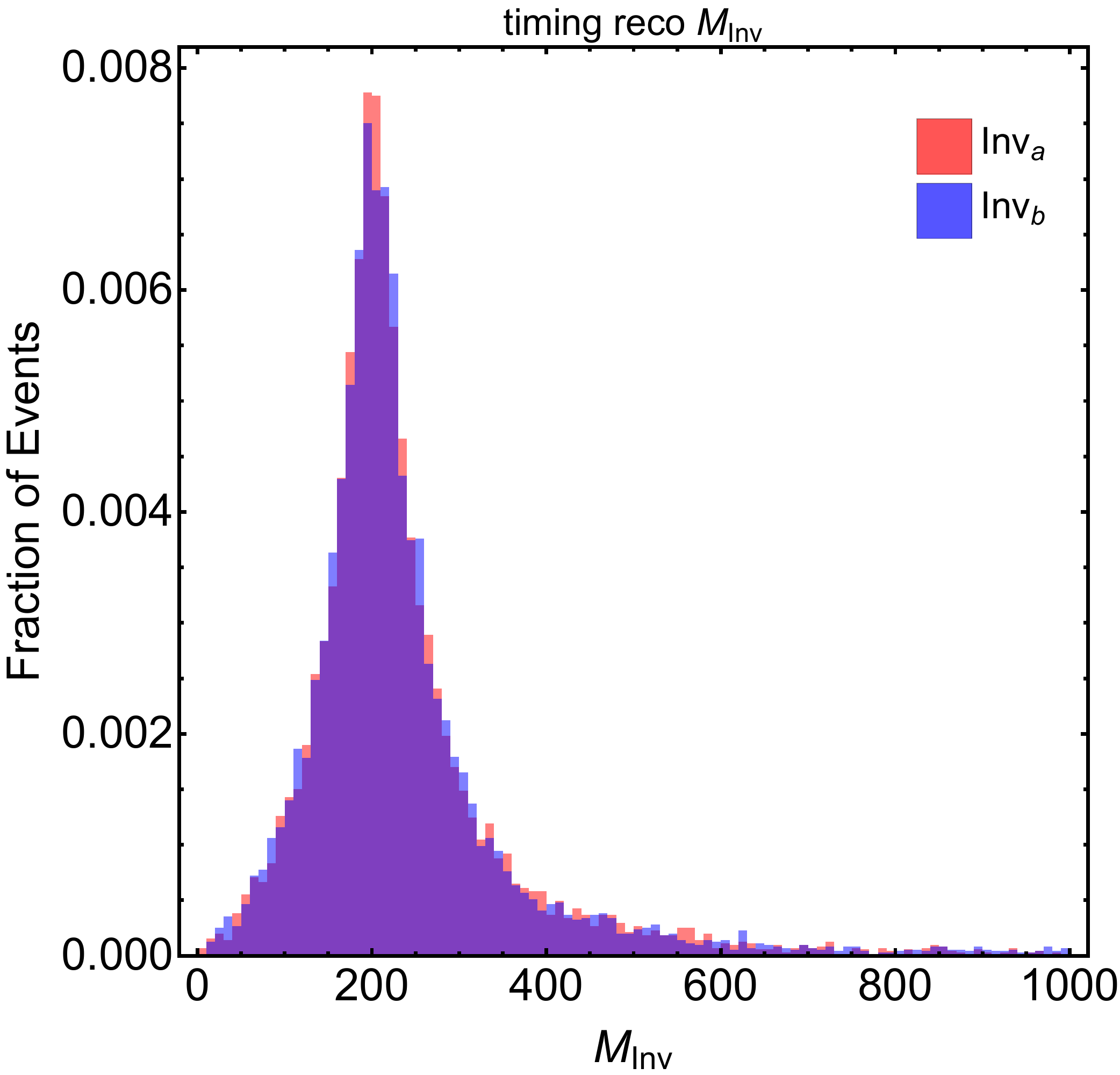}
    \includegraphics[width=0.3\textwidth]{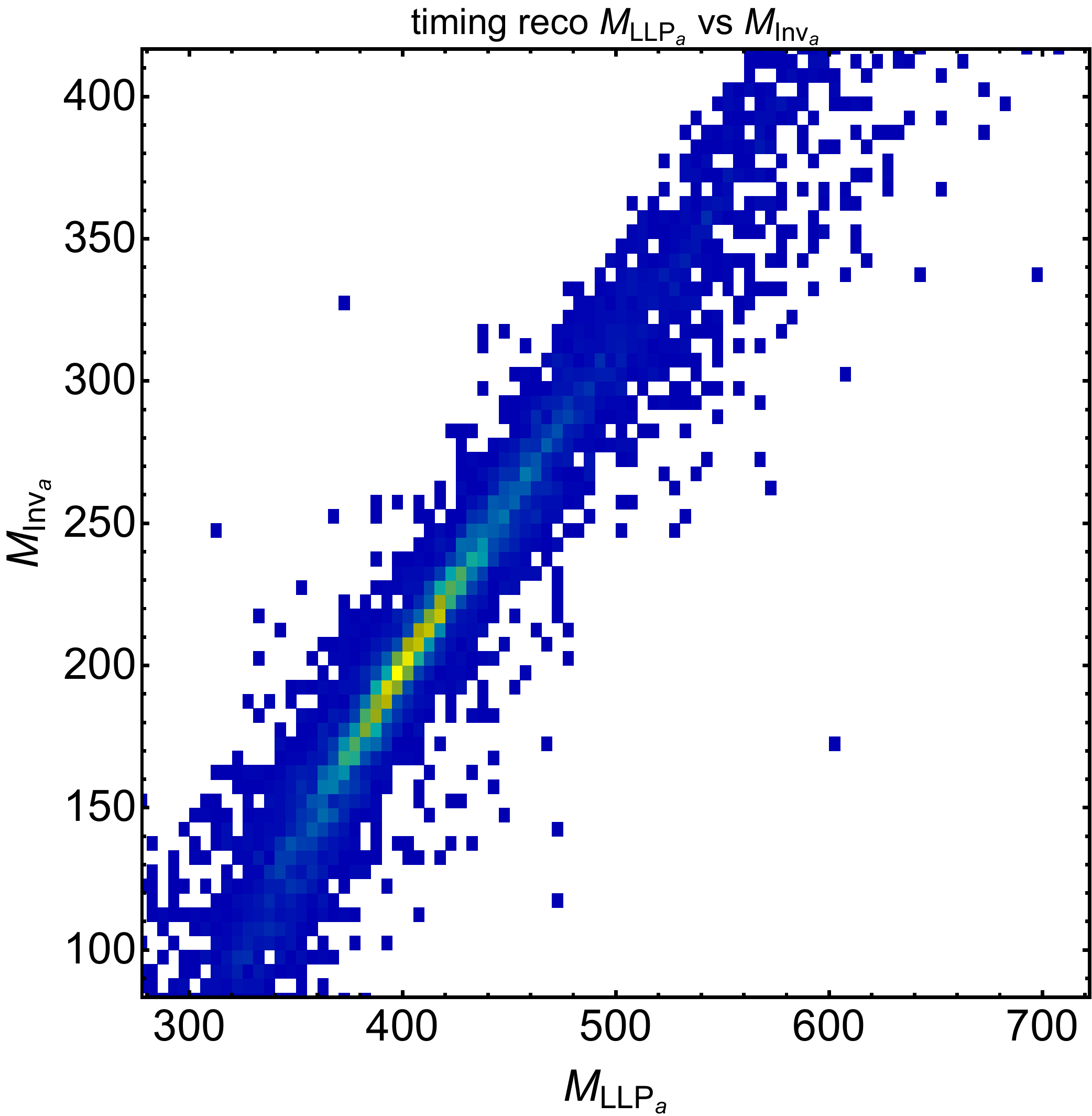}
    \includegraphics[width=0.32\textwidth]{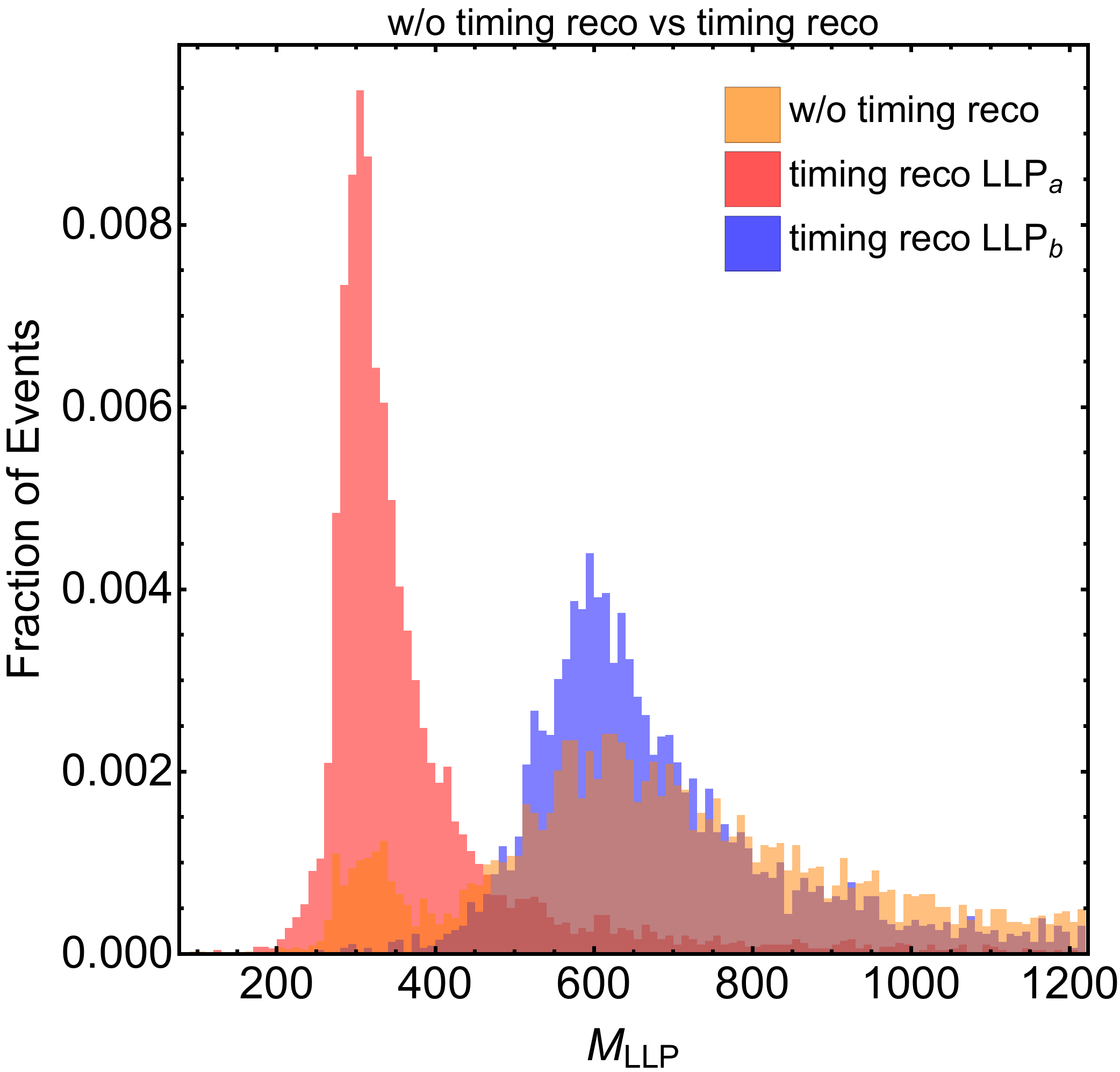}
    \includegraphics[width=0.32\textwidth]{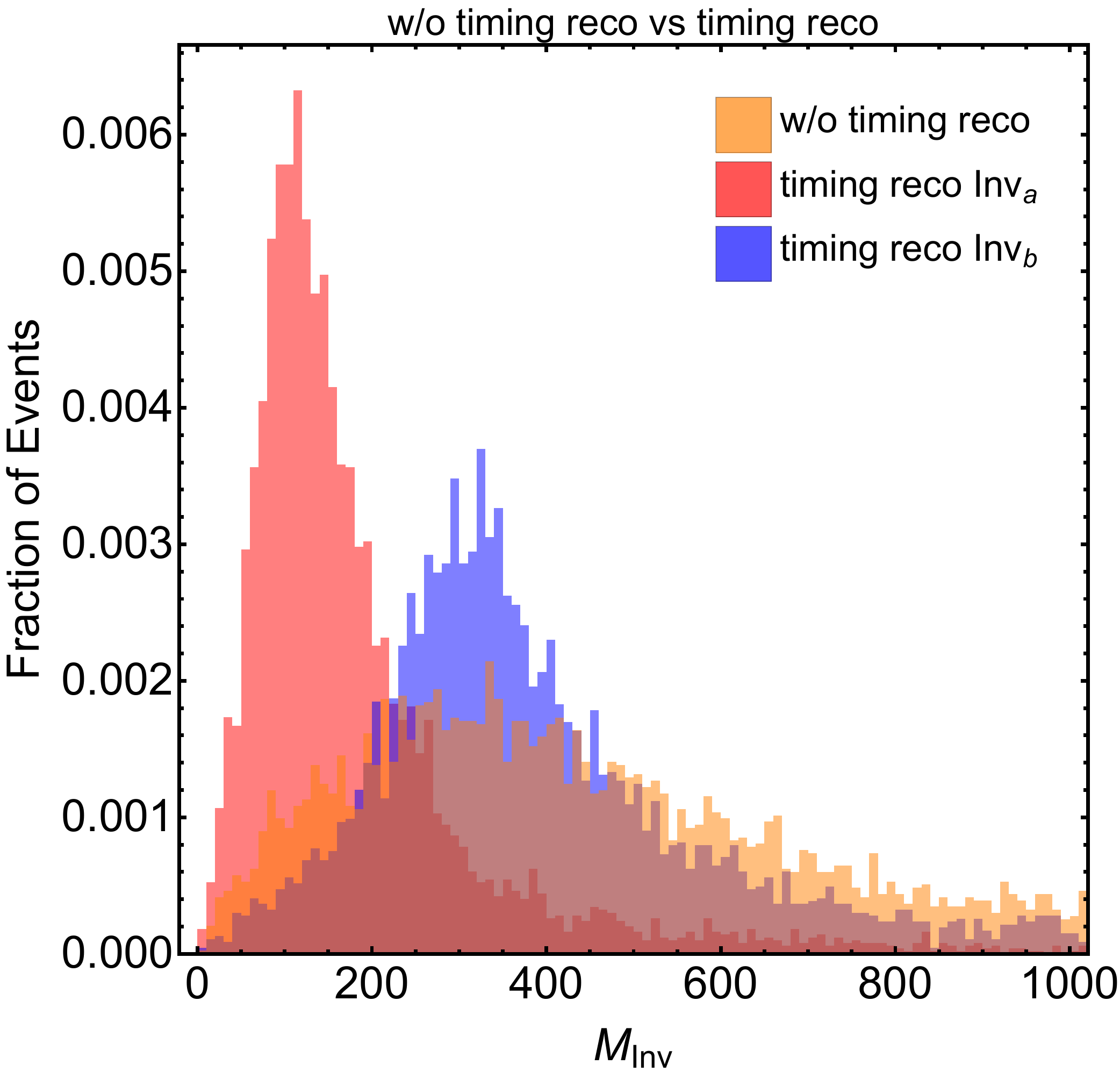}
    \includegraphics[width=0.3\textwidth]{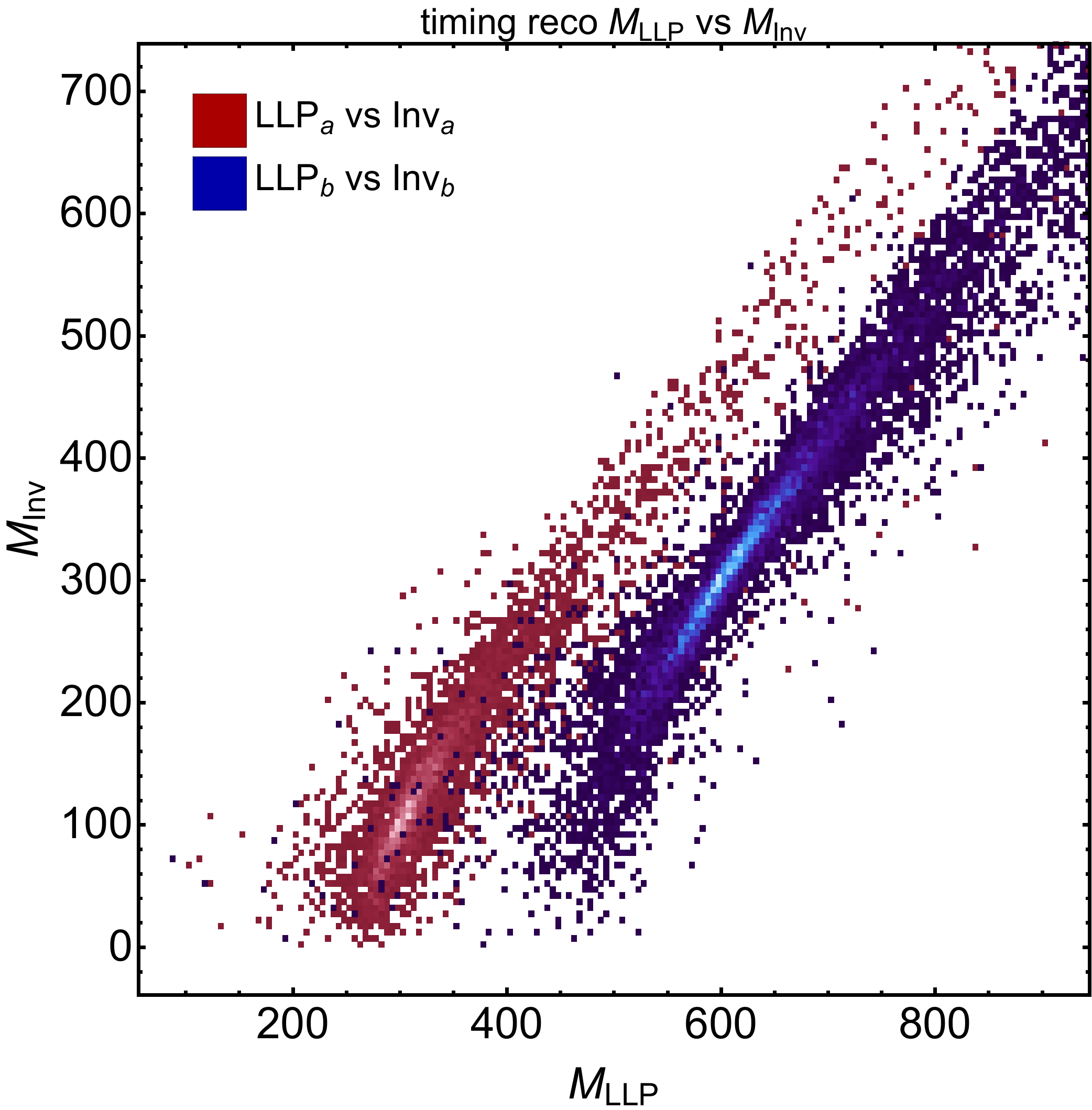}    
    \caption{Mass reconstruction without timing information (top) and with timing information (bottom) for $M_{\rm LLP_{a}}=M_{\rm LLP_{b}}= 400$ GeV,  $M_{\rm Inv_{a}}=M_{\rm Inv_{b}}= 200$ GeV. (Left) LLP mass reconstruction; (Center) Invisible particle mass reconstruction. Red and blue color indicate each decay chain; (Right) Invisible mass vs. LLP mass distributions.}
    \label{fig:case1_fig_M_p_o_r_o}
\end{figure}

{\it Assuming} symmetry of the LLP decays, the reconstructed LLP and invisible particle mass distributions calculated using the ``w/o timing reconstruction" method are shown in the Fig.~\ref{fig:case1_fig_M_p_o_r_o}. Clean peaks are observed in the distributions of  $M_{\rm LLP }$ and $M_{\rm Inv}$ at the true values.

Appealing to additional timing information from the reconstructed vertices in these events, the masses $M_{\rm LLP_{a}}$ and $M_{\rm LLP_{b}}$ can be calculated in the same simulated events without {\it a priori} assumptions about LLP and invisible particle mass relations, with the results shown in Fig.~\ref{fig:case1_fig_M_p_o_r_o}.  The clear peaks near the true $M_{\rm LLP}$ and $M_{\rm Inv}$ values are prominently seen for each decay chain independently. 

\subsubsection{$LLP_{a} \neq LLP_{b}$  and  $I_{a} \neq I_{b}$}

The two mass reconstruction approaches are also compared in a scenario with non-identical LLP decays, choosing $M_{\rm LLP_{a}} = 300$ GeV, $M_{\rm LLP_{b}} = 600$ GeV, $M_{\rm Inv_{a}} = 100$ GeV, and $M_{\rm Inv_{b}} = 300$ GeV, with $c\tau \approx \mathcal{O}(100)$mm.

\begin{figure}[!thb]
    \centering 
    \includegraphics[width=0.32\textwidth]{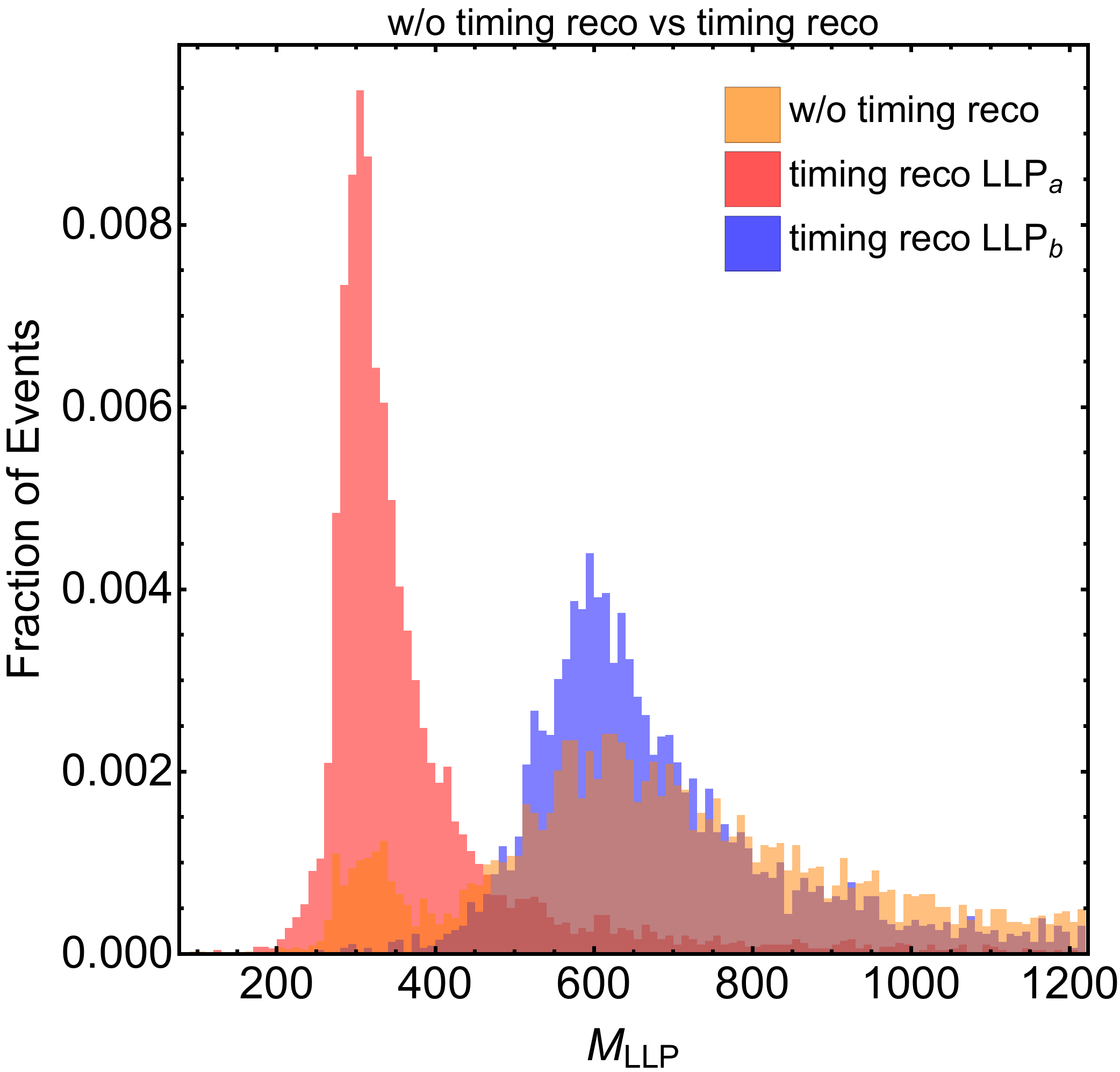}
    \includegraphics[width=0.32\textwidth]{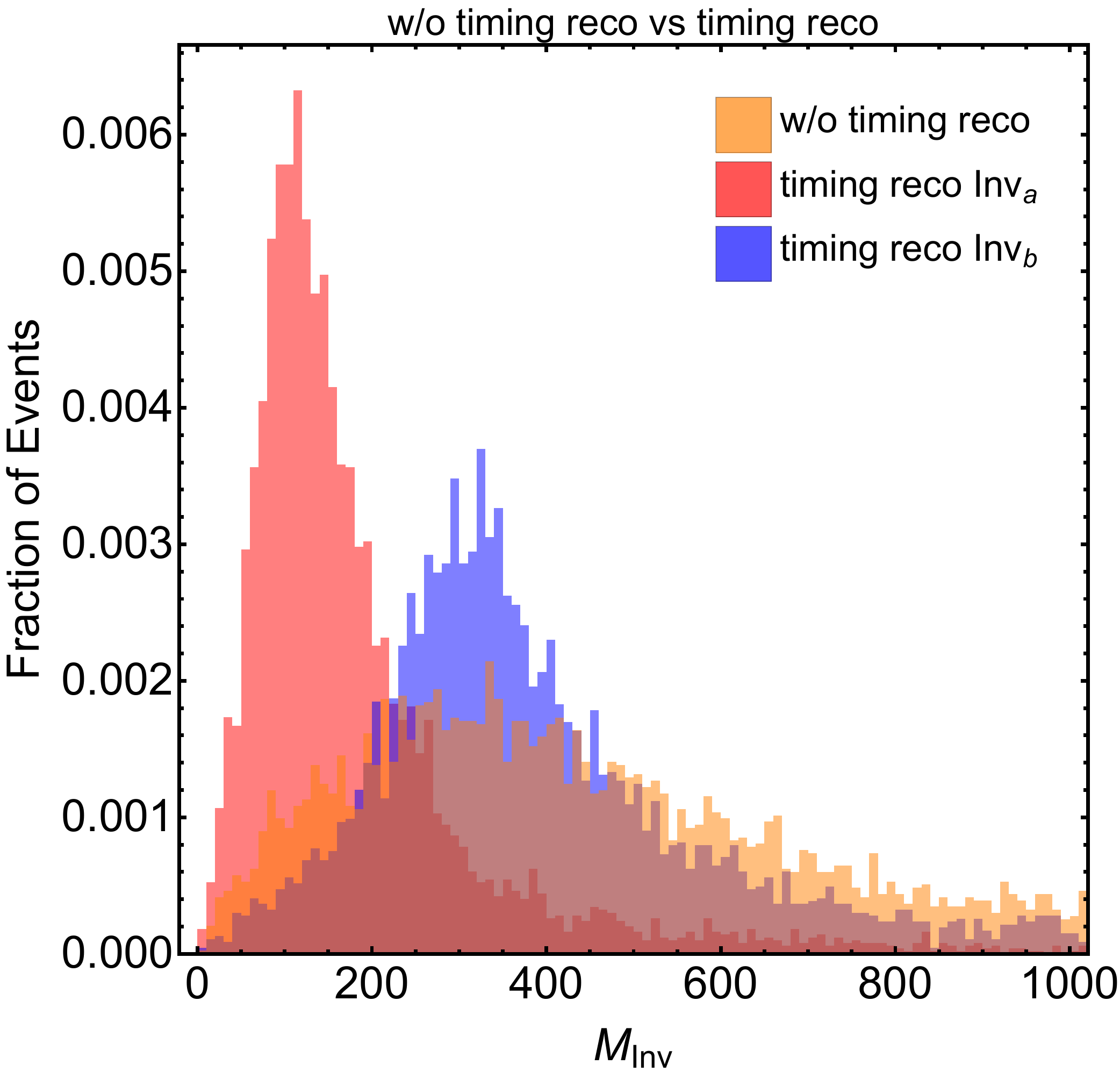}
    \includegraphics[width=0.3\textwidth]{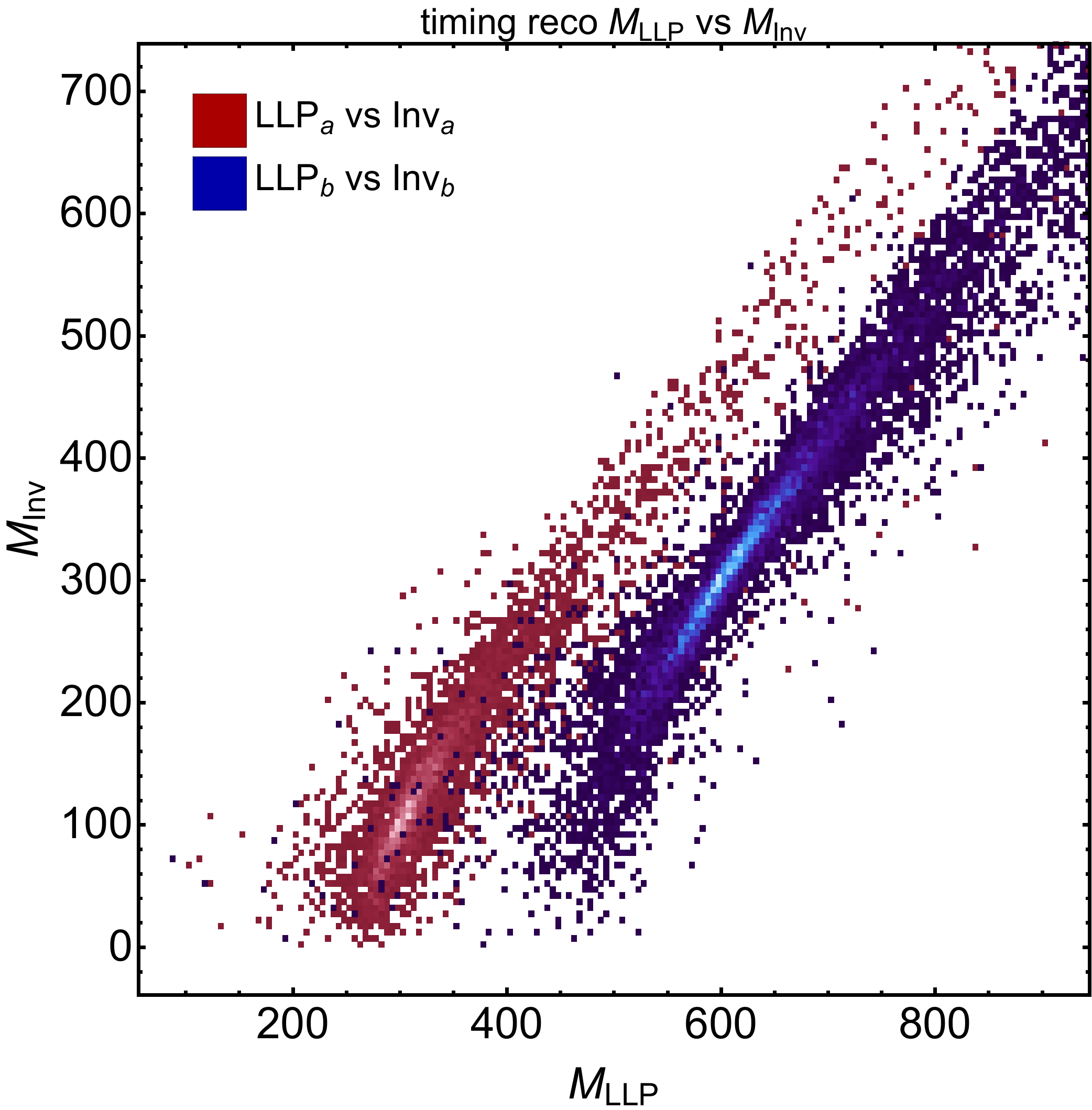}    
    \caption{Comparison with and without timing reconstruction. (Left) LLP mass reconstruction; (Center) Invisivle particle mass reconstruction. Red and blue color indicate each decay chain of timing reconstruction result and Orange color is without timing reconstruction result. (Right) invisible mass vs. LLP mass for each decay chain.}
    \label{fig:case2_fig_M_p_o_r_o_t_o_2}
\end{figure}

Masses calculated using both methods are shown in the Fig.~\ref{fig:case2_fig_M_p_o_r_o_t_o_2}. In this scenario, the assumption of symmetry between the LLP decays used in the ``w/o timing reconstruction" approach is violated, resulting in an incorrect determination of the masses. While the LLP and invisible particle masses can be determined independently for each decay, a strong correlation is observed between the measured masses in each decay. This behavior is further explored in Sec.~\ref{sec:timingreco}.

The masses ``measured'' from the simulated event distributions using both reconstruction methods are summarized in Table.~\ref{table:RecoSummary}. Included in the table are the  {\it reconstruction efficiencies} $\epsilon_{reco} = N_{reco}/N_{gen}$, corresponding to events with only complex (unphysical) solutions to the kinematic constraints due to imperfect detector resolution and, in the case of ``w/o timing reconstruction", incorrect assumptions. We observe that these methods can successfully infer the masses of LLPs and dark matter particles at HL-LHC.  

\begin{table}[h]
\centering
\begin{tabular}{ccccccc}
\hline 
&  & $m_{LLP_{a}}$ & $m_{LLP_{b}}$ & $m_{I_{a}}$ & $m_{I_{b}}$  &$ \epsilon_{reco}$ \\ 
\hline \hline
Identical LLPs &  w/o timing& 401 & 401 & 202 & 202 & 0.86 \\ 

 & timing & 405 & 404 & 206 & 206 & 0.72 \\ 
\hline 
Non-identical LLPs & w/o timing & - & - & - & - & - \\ 

 & timing & 319 & 633 & 132 & 342 & 0.51 \\ 
\hline 
\end{tabular} 
\caption{Reconstructed masses and reconstruction efficiencies.}
\label{table:RecoSummary}
\end{table}

\subsection{Timing reconstruction of neutral LLP decays}
\label{sec:timingreco}

\begin{figure}[!thb]
    \centering 
    \includegraphics[width=0.33\textwidth]{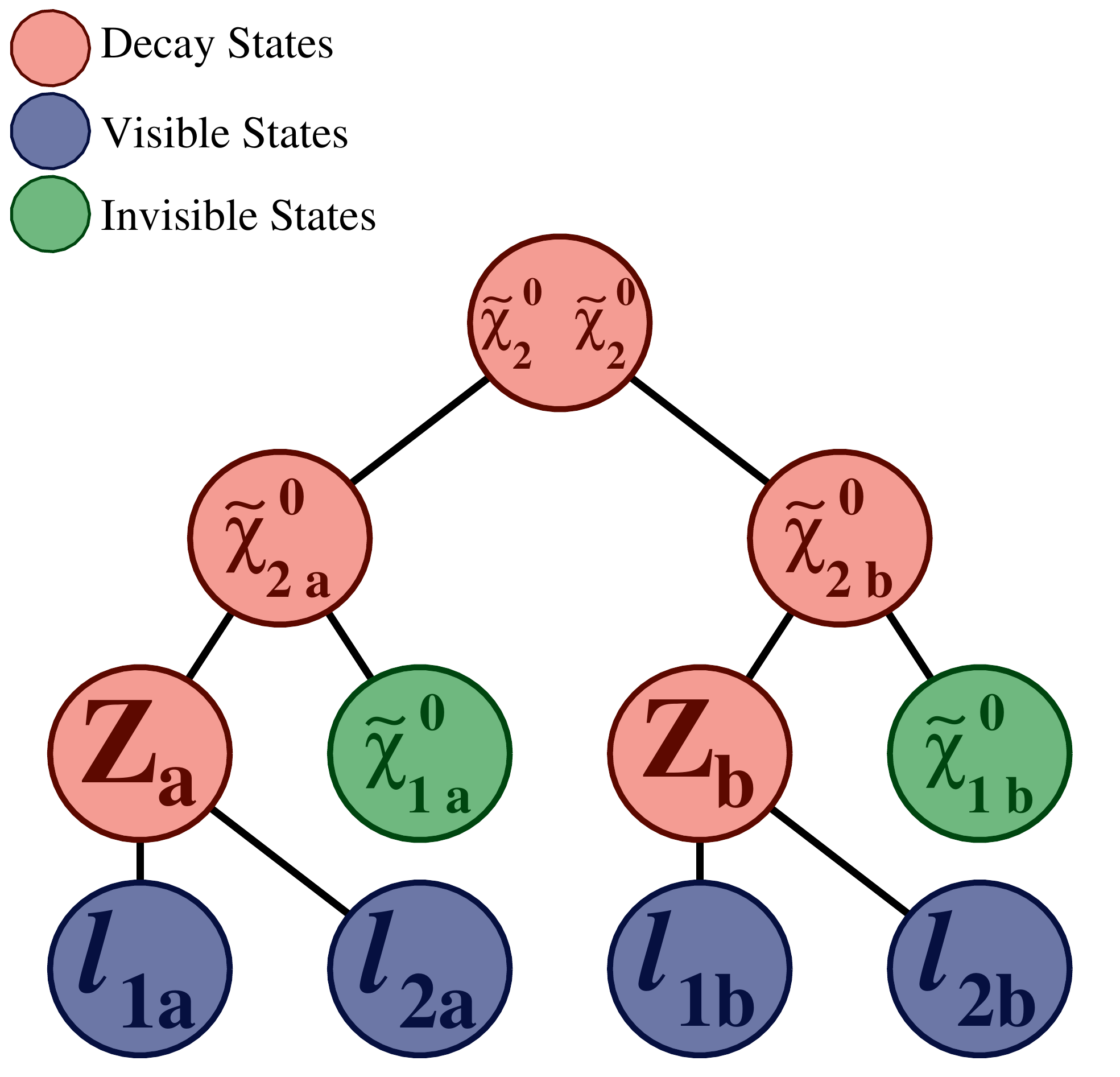}
    \hspace{1cm}
     \includegraphics[width=0.43\textwidth]{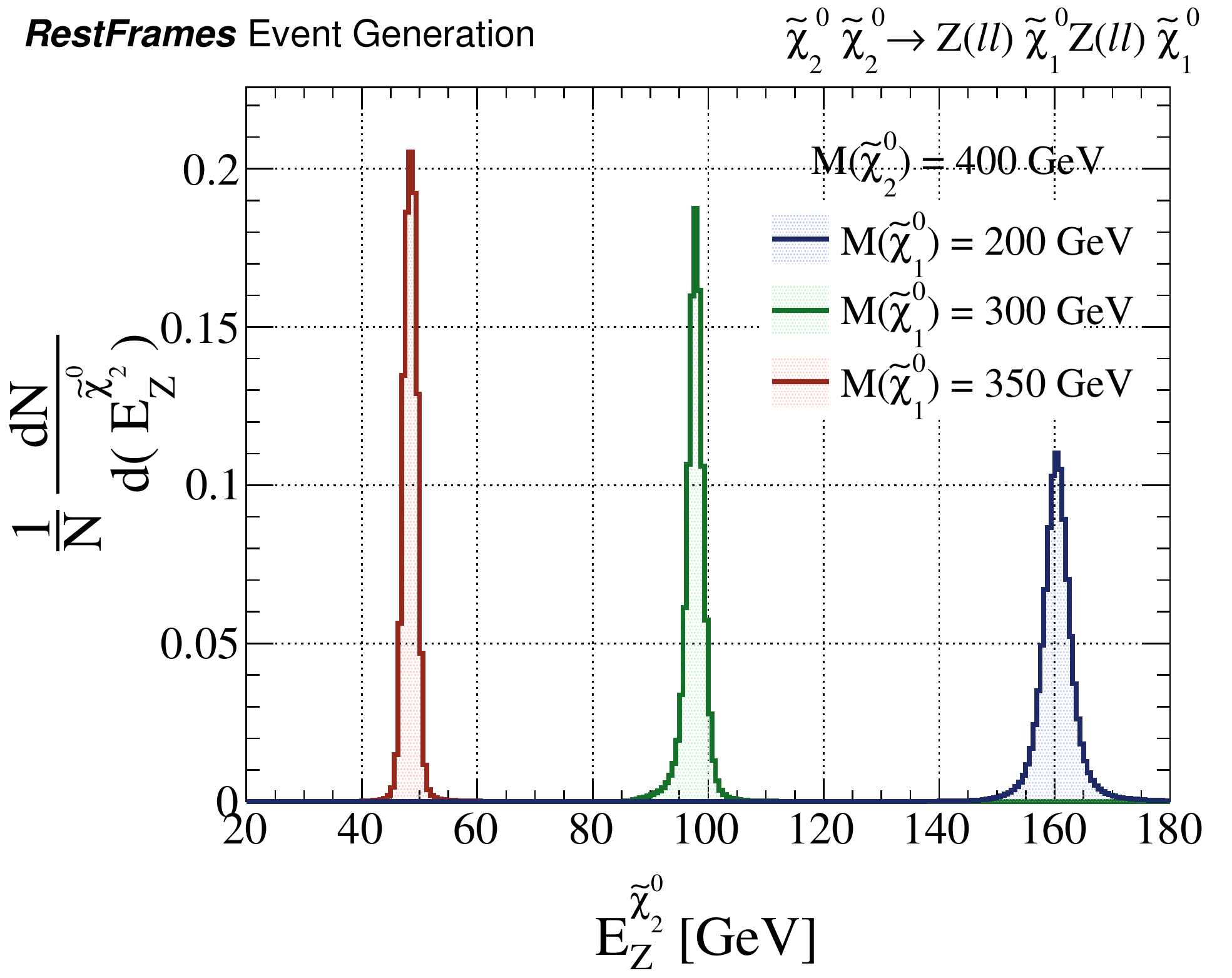}
    \caption{(Left) Decay topology for pair-production of long-lived neutralinos $\tilde{\chi}_{2}^{0}$, each decaying to a $Z$ boson and LSP $\tilde{\chi}_{1}^{0}$. (Right) Distribution of the energy of the $Z$ evaluated in the LLP rest frame for different LSP masses, assuming an LLP mass of 400 GeV.}
    \label{fig:decaytree}
\end{figure}

In order to evaluate how the precision of timing-assisted LLP decay reconstruction depends on the lifetimes and masses of particles appearing in these decays we consider a SUSY scenario with long-lived neutralinos, $\tilde{\chi}_{2}^{0}$. We assume that these LLPs are pair-produced in events, each decaying to a (possibly off-shell) $Z$ boson, which in turn decays to leptons, and an invisible lightest neutralino, $\tilde{\chi}_{1}^{0}$. The decays in this process are illustrated in Fig.~\ref{fig:decaytree}. 

To permit fast event simulation for varying masses and lifetimes, a two-step procedure is employed, whereby the kinematics of the di-LLP system is first modeled using \texttt{MG5aMC}~\cite{Alwall:2014hca}, including the simulation of up to two associated partons for this process for $\tilde{\chi}_{2}^{0}$ masses between 100 GeV and 1 TeV. The hadronization and decays of these events is then simulated with \texttt{Pythia8}~\cite{Sjostrand:2006za, Sjostrand:2007gs,Sjostrand:2014zea} (assuming zero LLP lifetime) and reconstructed using an emulation of the CMS HL-LHC detector in Delphes~\cite{deFavereau:2013fsa} with 200 pile-up interactions. These reconstructed events are used to develop a fast, parameterized model of the detector resolution for reconstructing lepton momentum and missing transverse energy. For the latter, particular care was taken to model the missing transverse energy resolution separately for the components parallel and perpendicular to the true missing momentum, respectively, while capturing correlations with the kinematics of the di-LLP system. 

Using the Madgraph-based model of the di-LLP system kinematics and fast HL-LHC detector-response parameterization, the remaining decays and kinematics of these events is simulated using the \texttt{RestFrames}~\cite{jigsaw} package, including simulation of displaced LLP decays. To incorporate the experimental effects of displaced vertex measurements, a 30 ps time resolution for individual charged particles was assumed unless otherwise noted, far in excess of the corresponding spatial vertex uncertainties, which were modeled to be less than 100 $\mu$m.

As described by Eq.~\ref{eq:EVLLPa}, the measured LLP velocities can each be used to evaluate the energy of the visible systems (here corresponding to the reconstructed $Z$ bosons) in their respective LLP rest frames, a quantity sensitive to the LLP and LSP masses in these events. Distributions of this reconstructed variable are shown in Fig.~\ref{fig:decaytree}, with narrow peaks at the combination of true LLP and LSP masses described by Eq.~\ref{eq:EVLLPb}. While such an observable results in a striking signature for these LLP signals, it is sensitive to the mass difference between the LLP and LSP in these decays rather than absolute masses. This ambiguity can be resolved by incorporating the measured missing transverse energy in the ``with timing'' reconstruction method introduced in this paper. 

\subsubsection{Mass reconstruction performance}

\begin{figure}[!thb]
    \centering 
    \includegraphics[width=0.48\textwidth]{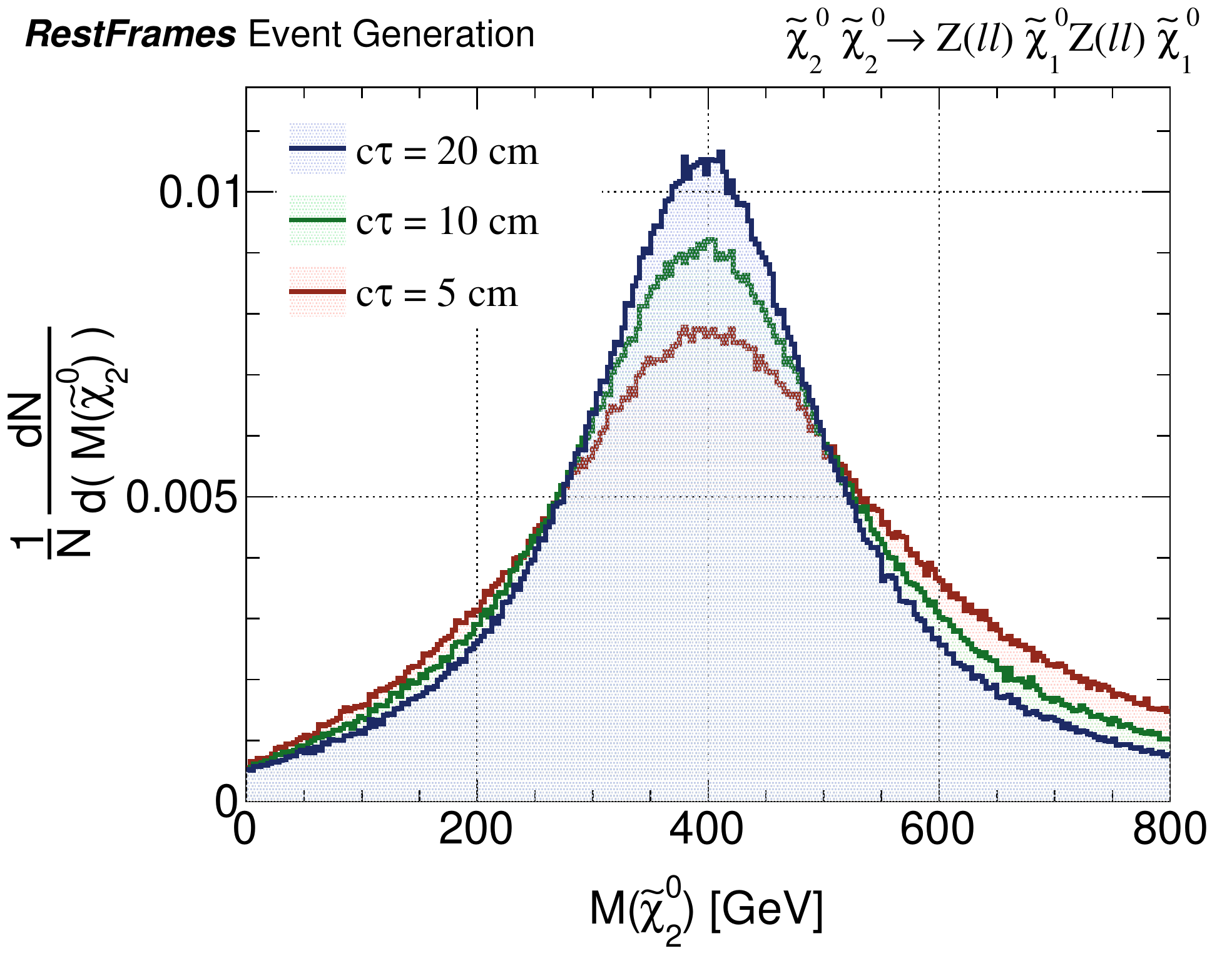}
    \includegraphics[width=0.48\textwidth]{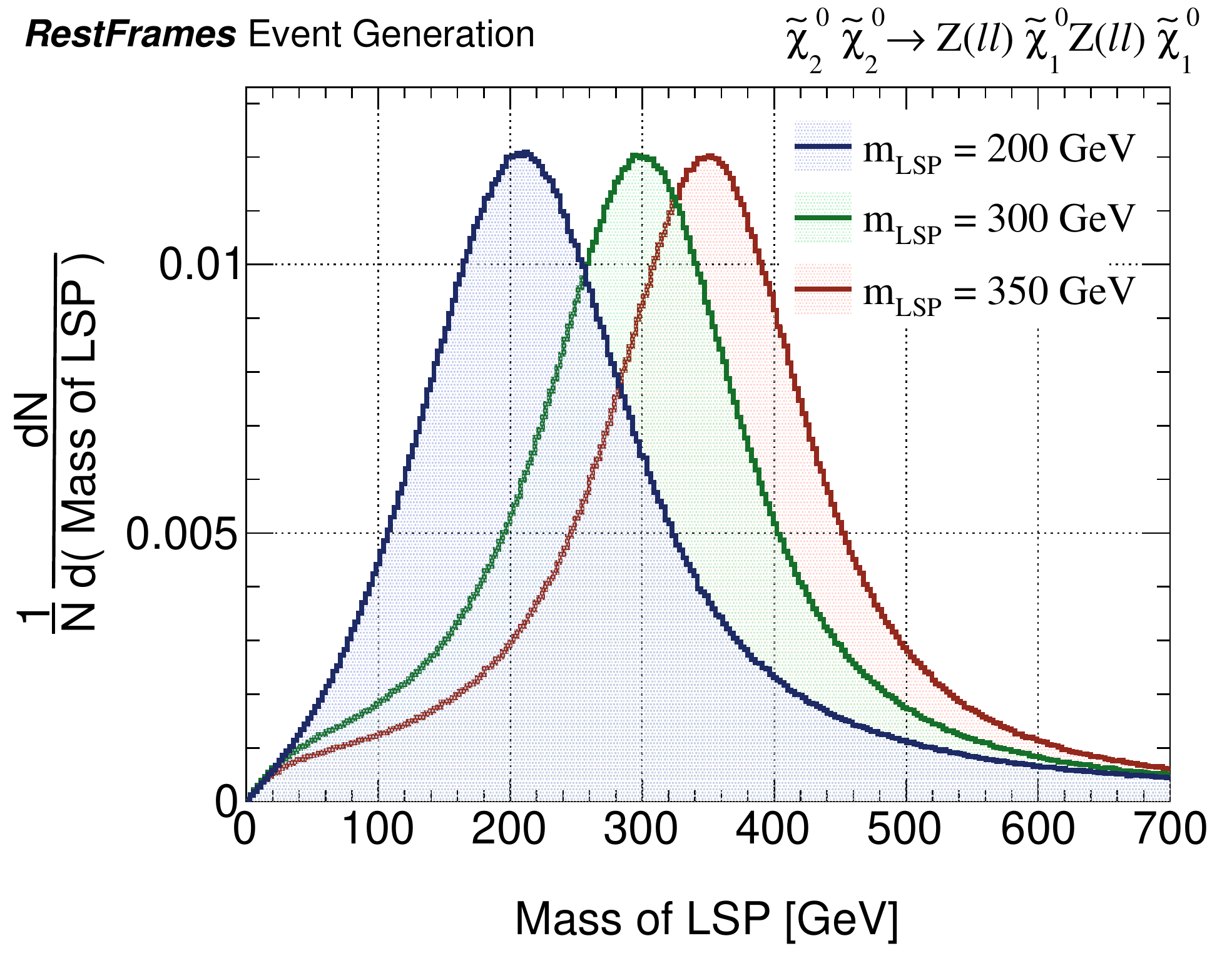}
    \caption{Distributions of reconstructed masses in simulated events. (Left) Reconstructed $M_{LLP}$ mass for different neutralino lifetimes. (Right) Reconstructed LSP mass for different true masses. An LLP mass of 400 GeV and lifetime of 20 cm is assumed when simulating these events unless otherwise noted.}
    \label{fig:MLLP}
\end{figure}

The distributions of reconstructed LLP and LSP masses for pair-production of 400 GeV long-lived $\tilde{\chi}_{2}^{0}$'s are shown in Fig.~\ref{fig:MLLP}. The mode for each of these reconstructed mass distributions corresponds closely to the true value, with resolution degrading with decreasing LLP lifetime. In general, these resolutions are worse relative to the visible energy observable shown in Fig.~\ref{decaytree} due to the dependence of calculated absolute masses on reconstructed missing transverse momentum.

\begin{figure}[!thb]
    \centering 
    \includegraphics[width=0.325\textwidth]{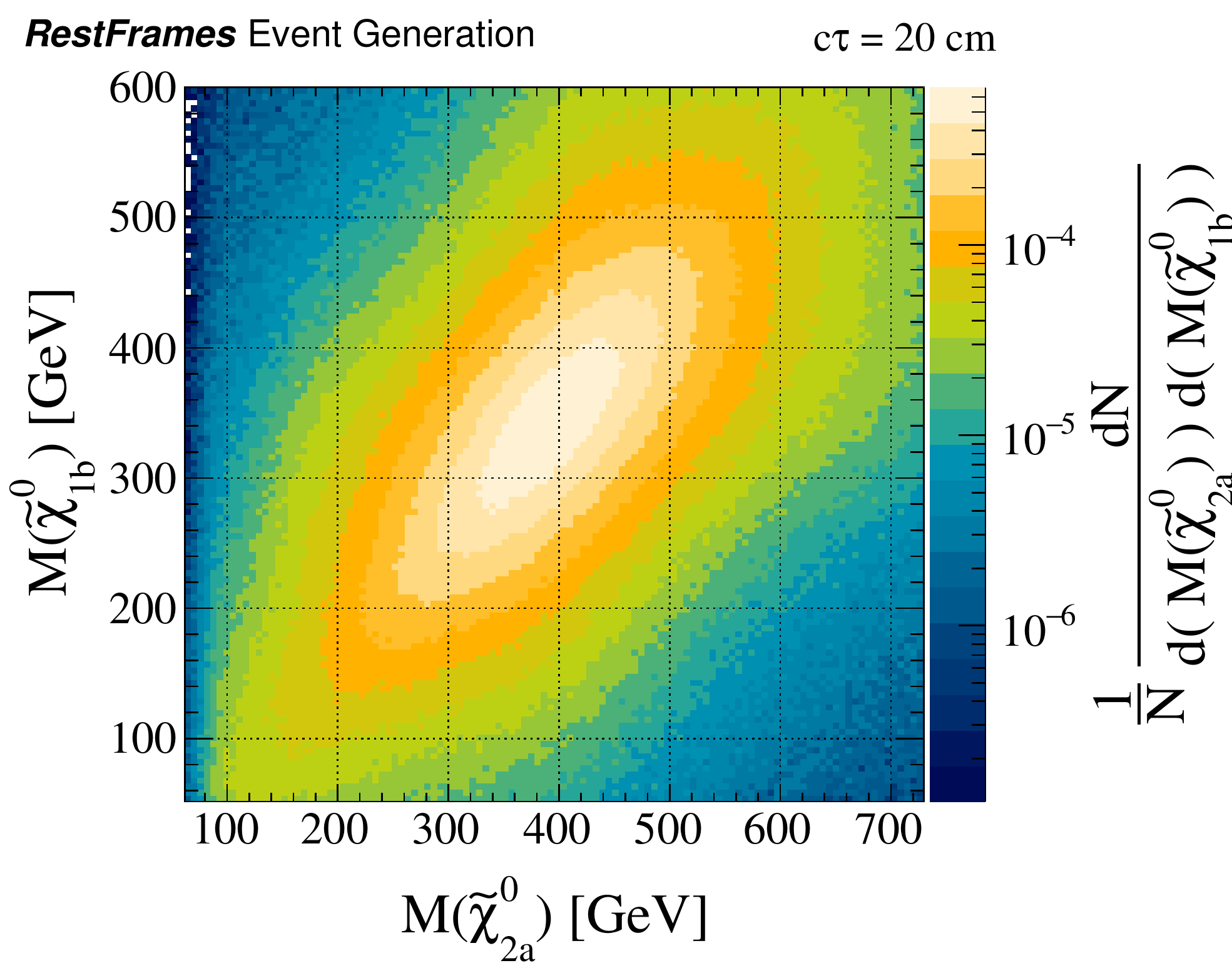}
     \includegraphics[width=0.325\textwidth]{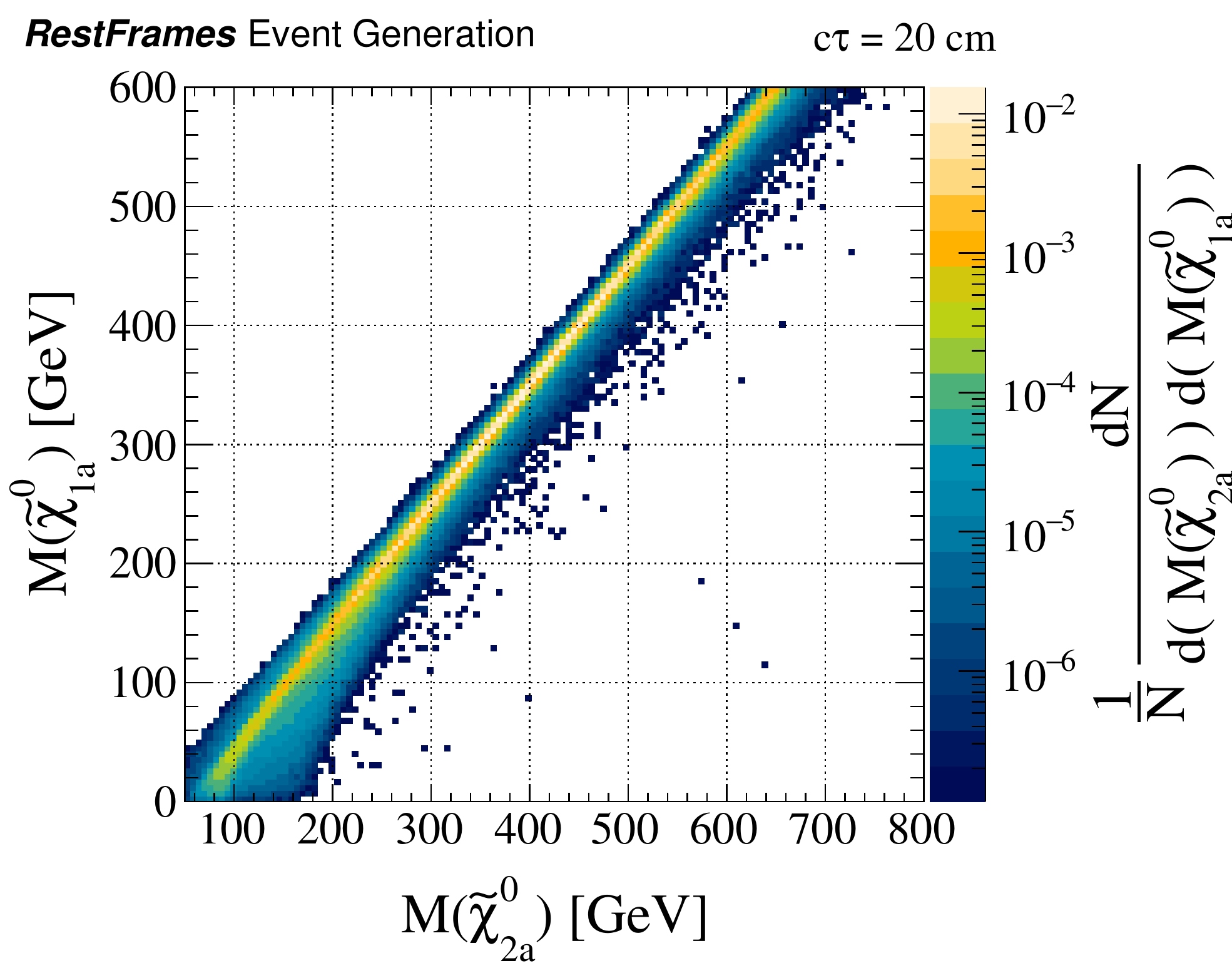}
     \includegraphics[width=0.325\textwidth]{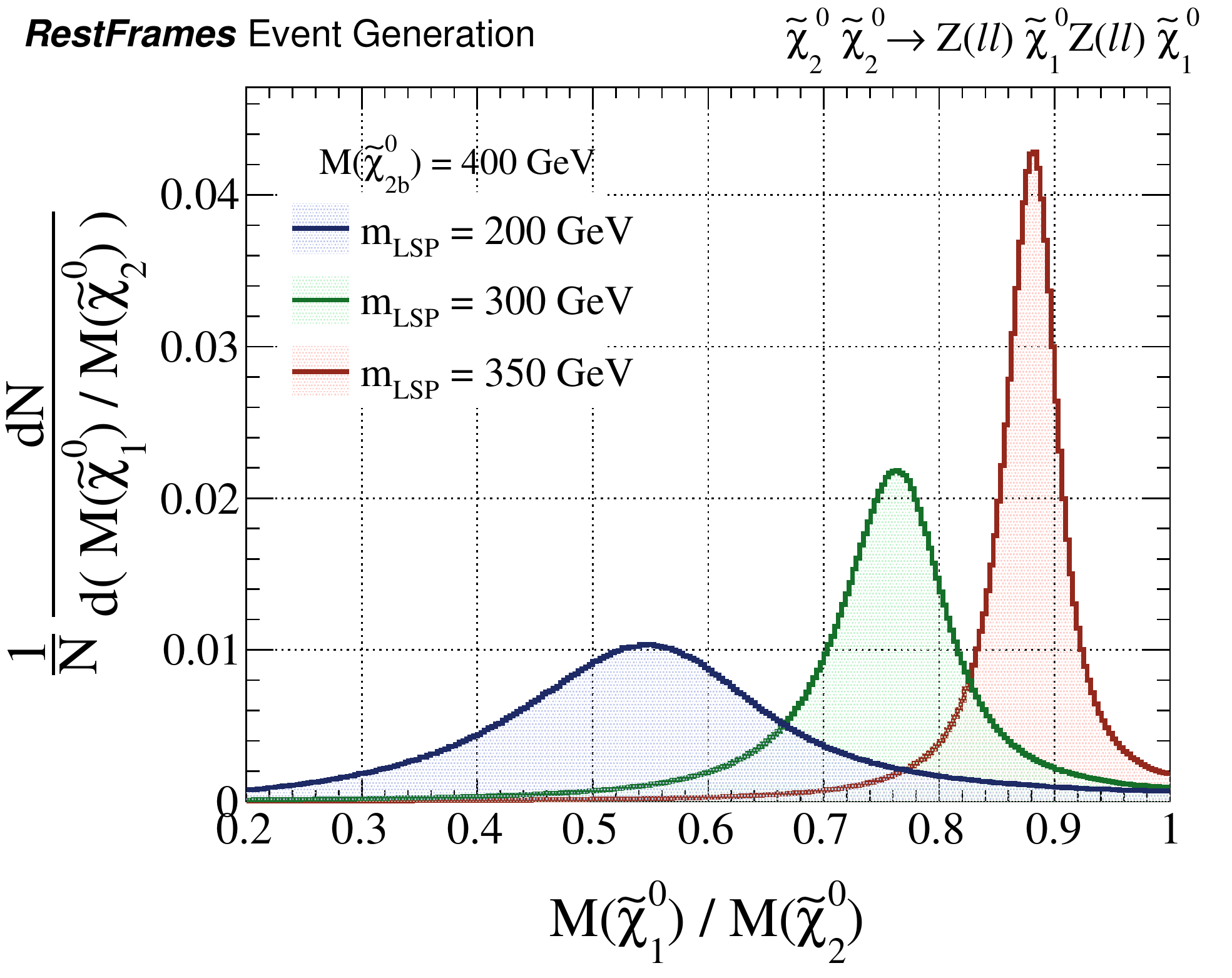}
    \caption{(Left) Distribution of reconstructed LSP mass vs. LLP mass, for separate decays in event. (Center) Distribution of reconstructed LSP mass vs. LLP mass, for the same decay. (Right) Distribution of reconstructed LSP and LLP mass ratio, for different LSP masses. An LLP mass of 400 GeV, LSP mass of 350 GeV, and lifetime of 20 cm is assumed when simulating these events unless otherwise noted.}
    \label{fig:MLSP_v_MLLP}
\end{figure}

Masses of the LLP and LSP are calculated independently for each separate LLP decay, with the resulting mass estimators largely uncorrelated between each half of the event, as can be seen in Fig.~\ref{fig:MLSP_v_MLLP} when looking at the distribution of the reconstructed LSP mass vs. LLP mass for separate decays. When looking at the same distribution for masses in the {\it same} decay, we observe that the calculated LLP and LSP masses are strongly correlated. As shown in Fig.~\ref{fig:MLSP_v_MLLP} (right), the ratio of reconstructed LSP and LLP masses for a single decay, is well-resolved, and is largely independent of the individual absolute masses.

\begin{figure}[!thb]
    \centering 
    \includegraphics[width=0.49\textwidth]{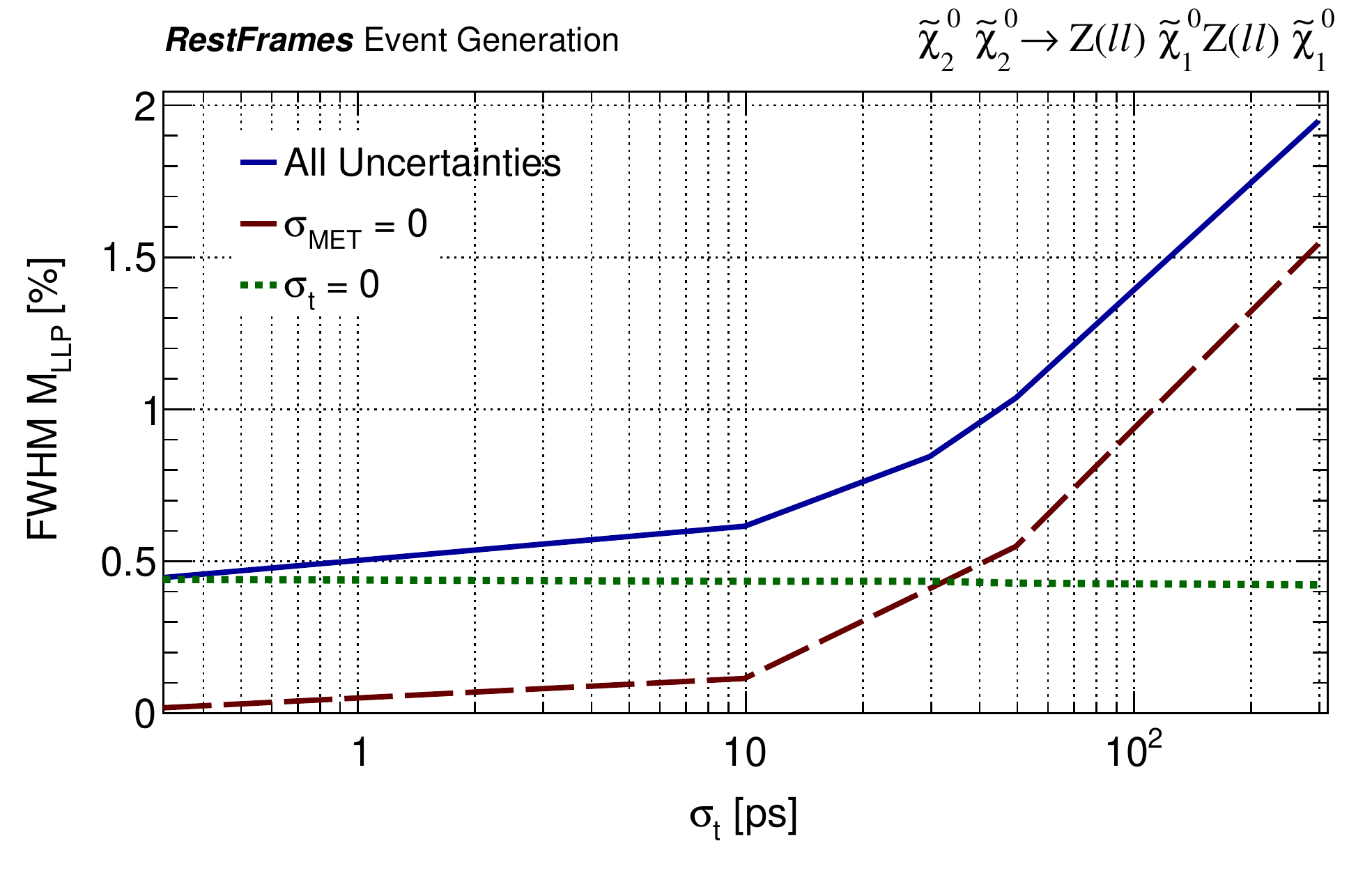}
     \includegraphics[width=0.49\textwidth]{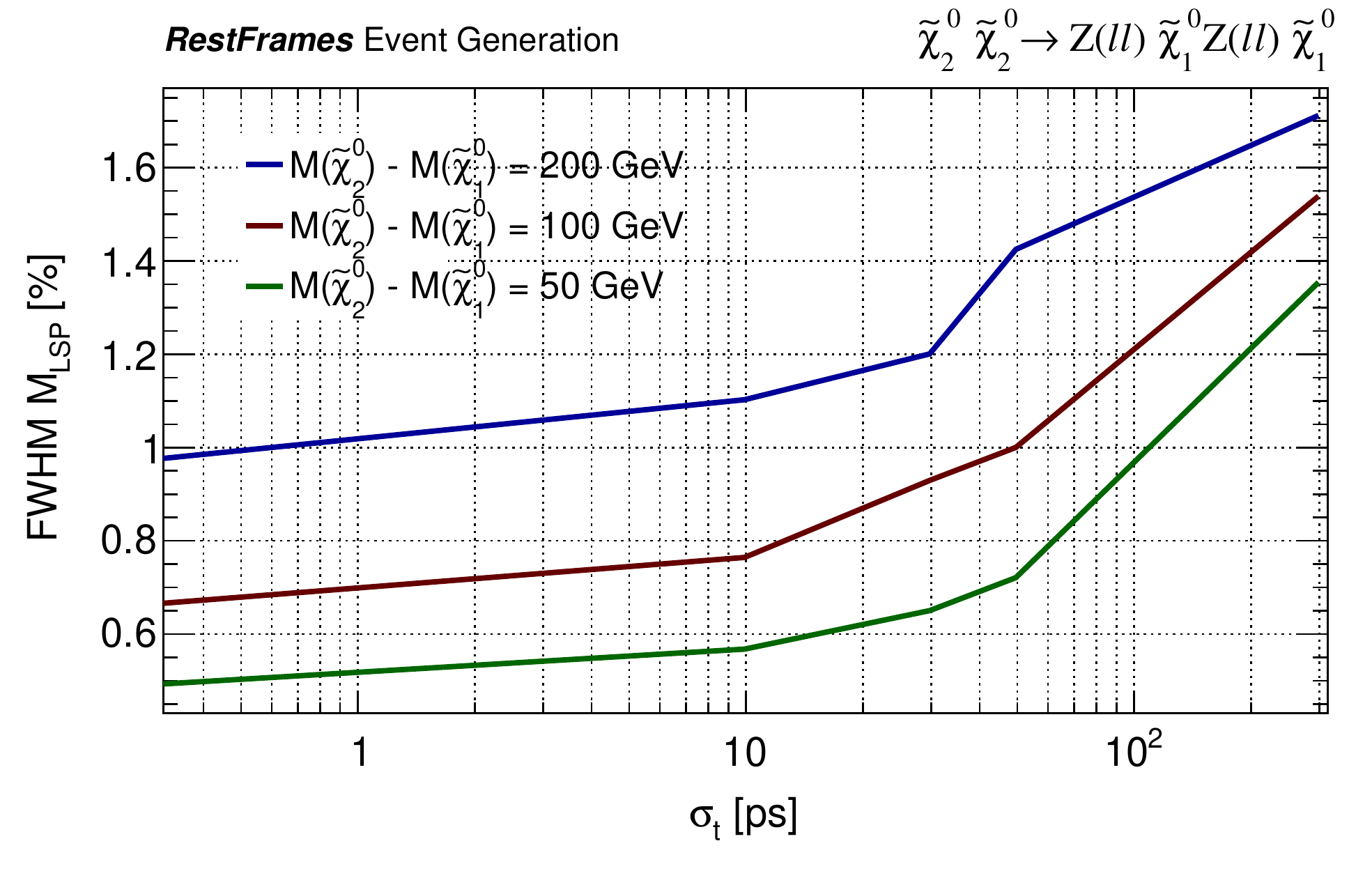}
    \caption{The full-width-half-maximum of reconstructed mass distributions as a function of timing detector single-track resolution. (Left) Width of the LLP mass distribution accounting for combinations of different experimental uncertainties. (Right) Width of the LSP mass distribution for different LLP/LSP mass splittings. An LLP mass of 400 GeV and lifetime of 20 cm is assumed when simulating these events unless otherwise noted.}
    \label{fig:Mres}
\end{figure}

The full-width-half-maximum (FWHM) of the reconstructed LLP and LSP mass distributions is evaluated as an estimator of the resolution of these distributions, and shown as a function of timing detector resolution in Fig.~\ref{fig:Mres}. For the LLP mass, experimental contributions to the mass resolution are isolated by disengaging different detector emulation effects. We observe that imperfect timing resolution and missing transverse energy reconstruction are by-far the dominant contributions to mass resolution, with spatial vertex resolution and lepton momentum reconstruction negligible in comparison. Coincidentally, the contributions from timing and missing transverse energy reconstruction to mass resolution have a similar magnitude for a time resolution corresponding to the proposed HL-LHC timing detectors. 

In the case of the reconstructed LSP mass, its resolution scales similarly to that of the LLP mass, with the addition of a factor approximately equal to the ratio of LLP and LSP masses, $m_{\rm LLP} / m_{\rm LSP}$. This results in the LSP mass resolution performing best in cases of increasingly compressed mass spectra, as seen in Fig.~\ref{fig:Mres}. Such compressed mass spectra appear naturally in many BSM scenarios, often as a reason for the long-lived nature of the decays. This makes this reconstruction approach for measuring invisible particle masses a precise tool for studying this otherwise difficult kinematic regime, where observables sensitive to mass-splittings can struggle to distinctively isolate signals. 

\subsubsection{Total event reconstruction}

\begin{figure}[!thb]
    \centering 
    \includegraphics[width=0.6\textwidth]{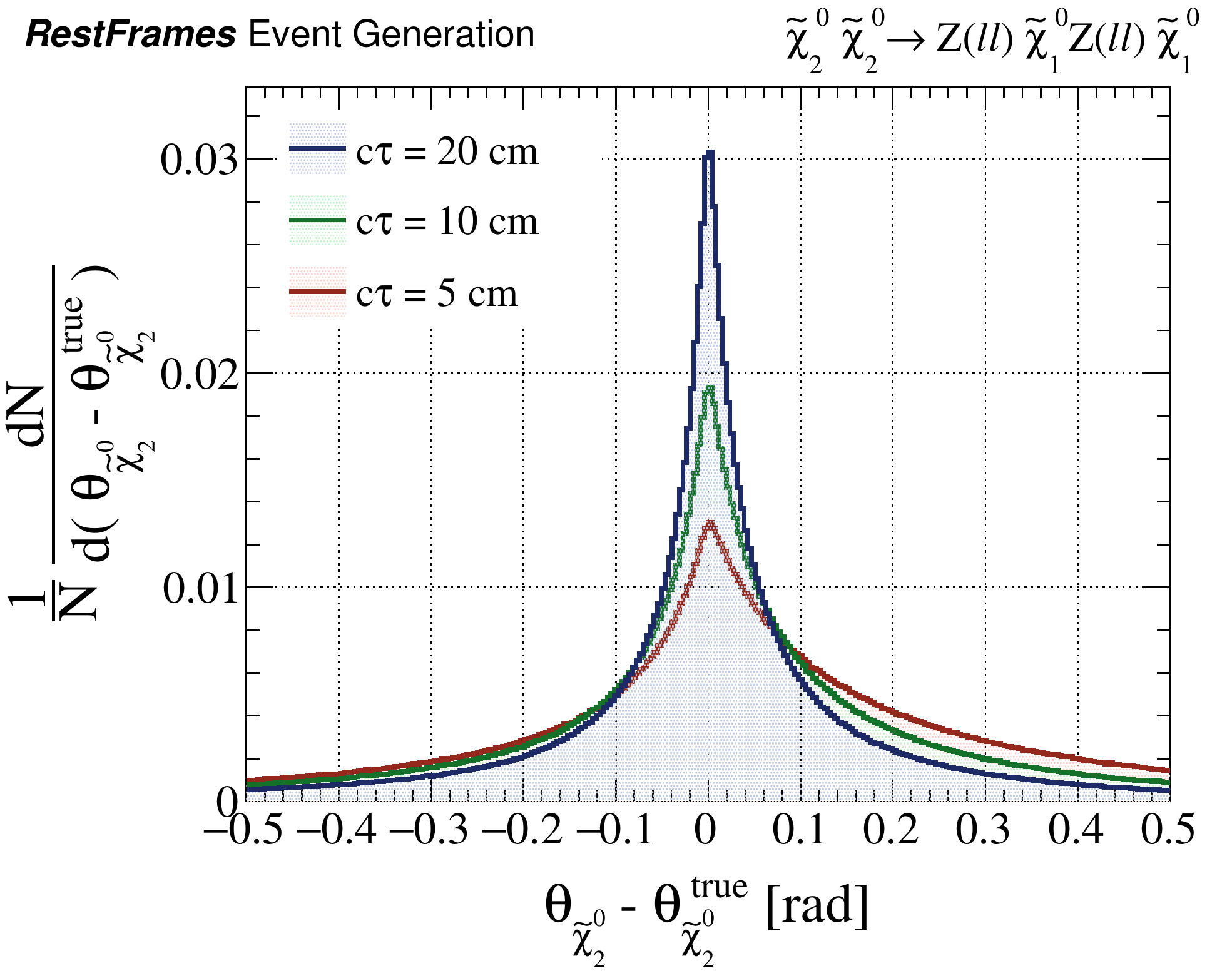}
    \caption{Distribution of different between true and reconstructed LLP decay angle for different neutralino lifetimes. }
    \label{fig:dtheta}
\end{figure}

While new particle masses are generally the observables of focus in searches for new particles, the proposed method of timing-assisted neutral LLP reconstruction allows for the kinematics of LLP decays to be reconstructed in their entirety, with other interesting variables available. One such observable is the decay angle of the LLP, which is sensitive to its spin. The distribution of this angle, compared to the true value it is estimating, is shown in Fig.~\ref{fig:dtheta}, where we observe that it is reconstructed with excellent resolution. Similarly, variables sensitive to the production mode of the LLPs can also be calculated.

\begin{figure}[!thb]
    \centering 
     \includegraphics[width=0.48\textwidth]{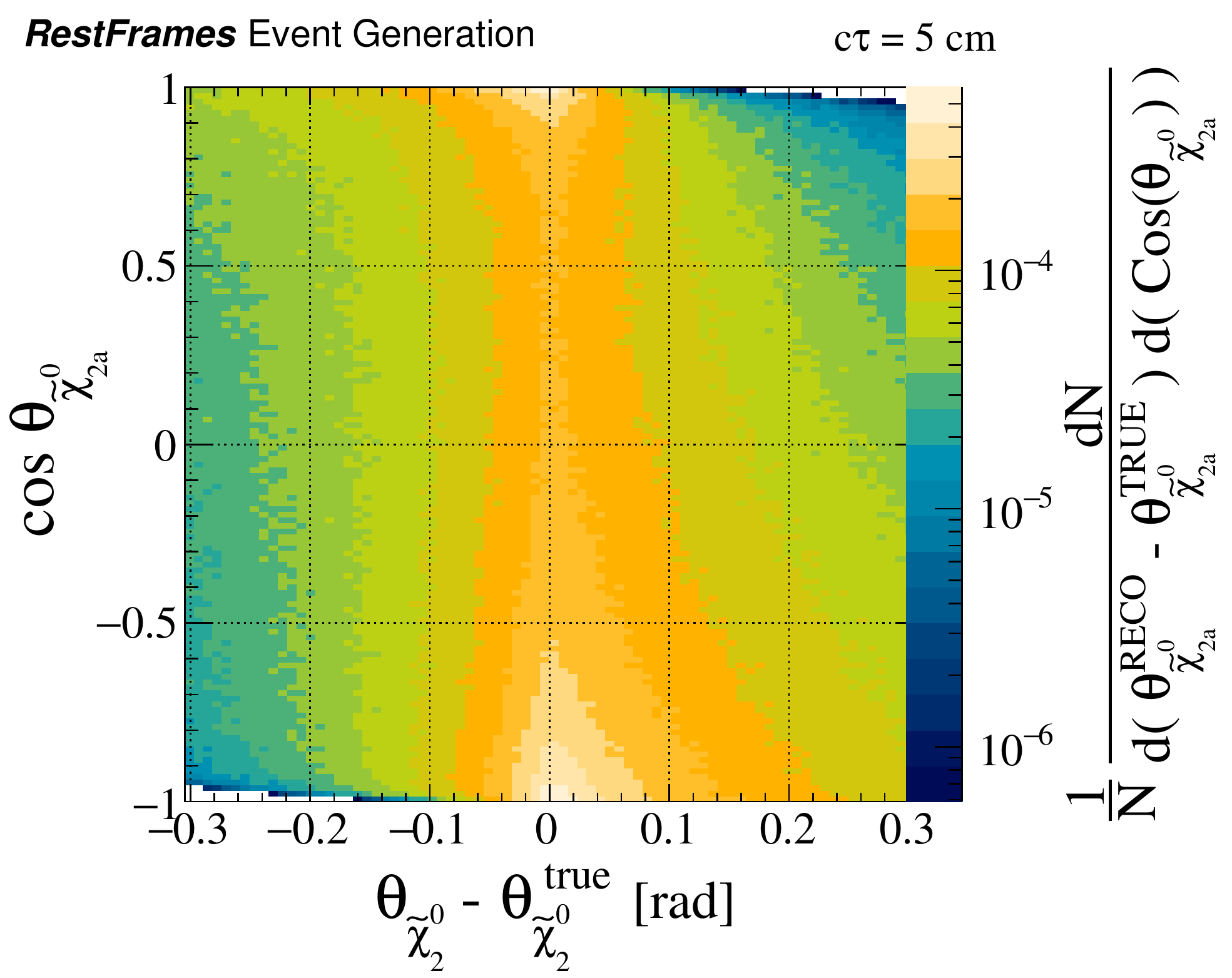}
      \includegraphics[width=0.48\textwidth]{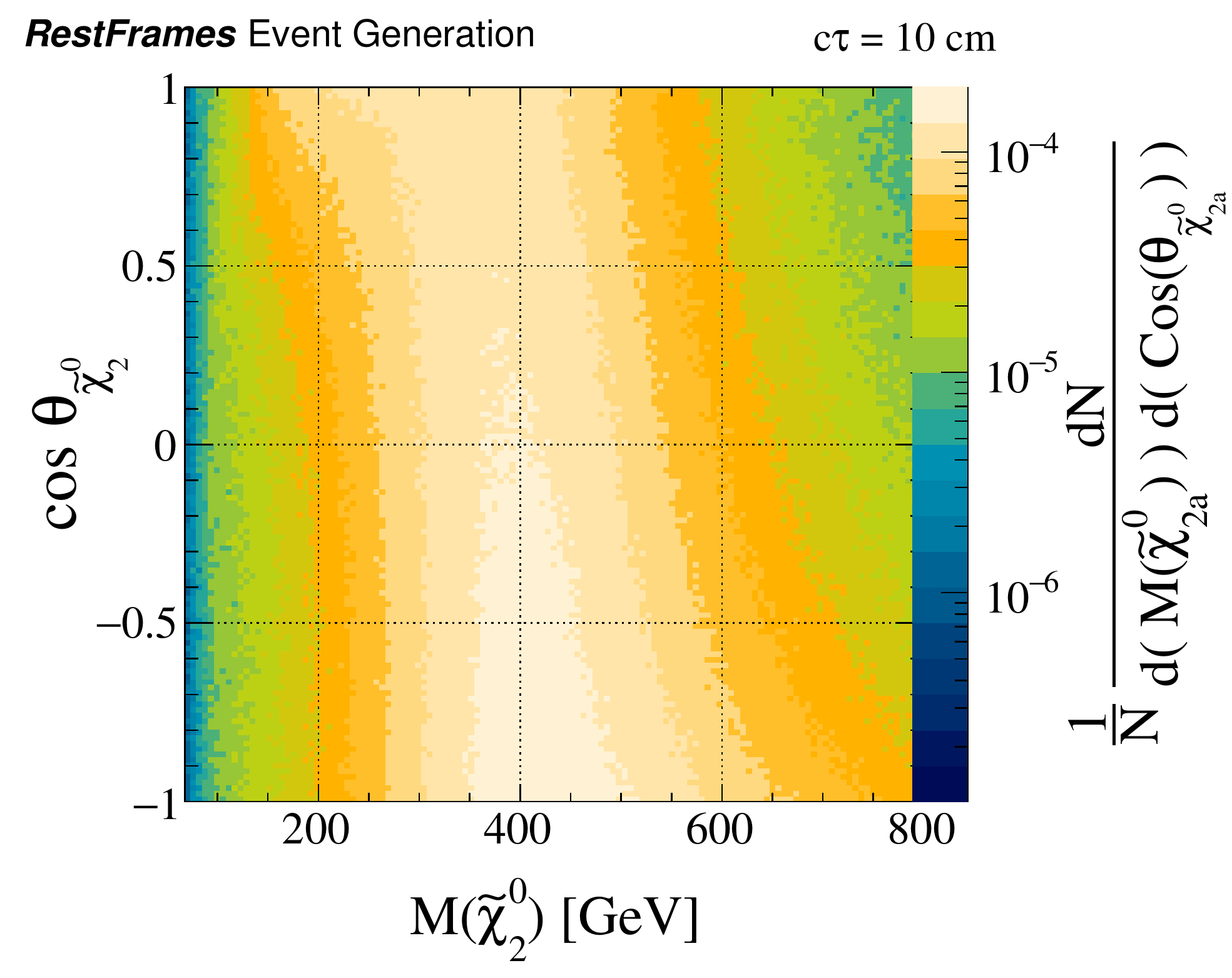}
    \caption{(Left) Cosine of the LLP decay angle as a function of reconstructed decay angle error. (Right) Cosine of the LLP decay angle as a function of reconstructed LLP mass. An LLP mass of 400 GeV is assumed when simulating these events.}
    \label{fig:costheta}
\end{figure}

The LLP decay angle is not only an interesting observable in what it could tell us about new physics, but also for practical reasons related to experimental effects in reconstruction. In Fig.~\ref{fig:costheta} we observe that deviations of the reconstructed decay angle from the true value are correlated with the cosine of that angle, such that the worst resolution corresponds to reconstructed values of ${\rm cos}~\theta_{\rm LLP}$ near -1 and 1. This behavior can be understood by considering the explicit formulation of this decay angle: When the LLP velocity is mis-measured due to imperfect timing resolution, the magnitude of the boost from the lab frame to the LLP rest frame is either over- or under-estimated. This induces an artificial correlation between the accuracy of the velocity measurement/boost magnitude and the measured decay angle, which is defined as the angle between the visible system momentum and boost direction in the LLP rest frame. Hence over- and under-boosts will preferentially align these axes.

The practical consequence of this observation is that events with reconstructed values of ${\rm cos}~\theta_{\rm LLP}$ near -1 and 1 are more likely to be poorly measured. This can be seen by looking at the correlation between ${\rm cos}~\theta_{\rm LLP}$ and the reconstructed LLP mass, shown in Fig.~\ref{fig:costheta}. The largest deviations from the true LLP mass, 400 GeV in this case, occur when $|{\rm cos}~\theta_{\rm LLP}| \sim 1$. The resolution of calculated masses and other observables can effectively be improved (at some cost in efficiency) by selectively removing events that are more likely to be poorly measured, without any prior knowledge of the true values of any new particle masses.

\section{Conclusion}
\label{sec:conclusion}

The precision timing detectors proposed by the CMS and ATLAS experiments at the HL-LHC have the potential to open an entirely new experimental window to the kinematics of new, long-lived particles. In particular, for the long lived particle (LLP) decaying to a dark matter particle (i.e. invisible particle) and a visible particle (i.e. the standard model particle(s)) the addition of timing information allows us to completely determine the masses of the LLP as well as the dark matter particle, information that is otherwise inaccessible. In this paper, we have developed  two novel reconstruction methods: (i) the first based on the precision displaced vertex measurement and applicable when the pair produced LLPs are identical and each of the LLP decays to the same dark matter particle. (ii) the second relying on the timing information of the long lived particle(s) and generally applicable to cases of two different LLPs decaying to different invisible particles. Evaluation of the expected performance of timing reconstruction of LLP kinematics indicates that this approach can accurately reconstruct the masses and kinematics of all the particles appearing in these events, invisible and long-lived, for a wide class of lifetimes and masses at the HL-LHC.

\acknowledgments
We thank K.C. Kong for useful discussion in the early stage of this study and also G. Cottin for communication. CR would like to thank his CMS colleagues for useful discussion in applications of precision timing measurements and their work towards creating a detector capable of realizing them.
This work was supported in part by the National Research Foundation of Korea (NRF) grant funded by the Korean government (MSIP)  (NRF- 2018R1A4A1025334 and 2019R1A2C1089334). A part of this work is done during SC's stay in CERN over the CERN-TH institute `Scale invariance in particle physics and cosmology' in January 2019. 

\appendix
\section{Determination of 3 momenta with displaced vertices} 
\label{appendix}

In this section, we will show that we can determine the 3-momenta of LLPs (but not the energies or equivalently masses of LLPs) even without timing information under the following conditions:
\begin{itemize}
	\item{\it Assumption-1:} We measure the displaced vertex of the LLP,
	\item{\it Assumption-2:} MET is only from $I_a$ and $I_b$,
	\item{\it Assumption-3:} We  fully reconstruct the 3-momentum of $V$ (with the known mass, $m_V$).
\end{itemize}

{\it Proof:}

We notice that the energy of ${\rm LLP}_{a}$ in lab frame is found to be related with 3-velocities, $\boldsymbol{\beta}_a$ and $\boldsymbol{\beta}_b$ as Eq.~\ref{eq:energy_LLP}. Let us derive this relation first,
Eq.~\ref{eq:llp2}
We can find the energy by cross producting $\boldsymbol{\beta}_{i, T}$ and dot producting to the beam axis $\hat{\boldsymbol{k}}$.
\begin{align}
E_{a} &= \left[ \frac{\boldsymbol{\beta}_{b,T}\times(\boldsymbol{p}_{I,T} + \boldsymbol{p}_{V_{a},T} + \boldsymbol{p}_{V_{b},T} )\cdot \hat{\boldsymbol{k}}}{\boldsymbol{\beta}_{b,T} \times \hat{\boldsymbol{\beta}}_{a,T} \cdot \hat{\boldsymbol{k}}}\right]   \nonumber \\
 &= \left[\frac{\boldsymbol{\beta}_{b}\times(\boldsymbol{p}_{T}^{miss} + \boldsymbol{p}_{V_{a}} + \boldsymbol{p}_{V_{b}} )\cdot \hat{\boldsymbol{k}}}{\boldsymbol{\beta}_{b} \times \boldsymbol{\beta}_{a} \cdot \hat{\boldsymbol{k}}} \right], \\
E_{b} &= \left[\frac{\boldsymbol{\beta}_{a}\times(\boldsymbol{p}_{T}^{miss} + \boldsymbol{p}_{V_{a}} + \boldsymbol{p}_{V_{b}} )\cdot \hat{\boldsymbol{k}}}{\boldsymbol{\beta}_{a} \times \boldsymbol{\beta}_{b} \cdot \hat{\boldsymbol{k}}} \right].
\end{align}
In the second line we have used the {\it Assumption-2} and the vector identity, $\hat{\boldsymbol{k}} \times \vec{V} \cdot \hat{\boldsymbol{k}}=0$ for an arbitrary vector $\vec{V}$ after decomposing the vectors into longitudinal ($\propto \hat{\boldsymbol{k}}$) and perpendicular components.

As the momentum is related with the energy by the relation $\boldsymbol{p}_{a} = E_{a}\boldsymbol{\beta}_{a}$, the 3-momenta of $LLP_a$ and similarly to the $LLP_b$ are obtained as
\begin{align}
\boldsymbol{p}_{a} &= \left[ \frac{\boldsymbol{\beta}_{b}\times(\boldsymbol{p}_{T}^{miss} + \boldsymbol{p}_{V_{a}} + \boldsymbol{p}_{V_{b}} )\cdot \hat{\boldsymbol{k}}}{\boldsymbol{\beta}_{b} \times\boldsymbol{\beta}_{a} \cdot \hat{\boldsymbol{k}}}\right]  \boldsymbol{\beta}_{a} \\
\boldsymbol{p}_{b} &= \left[ \frac{\boldsymbol{\beta}_{a}\times(\boldsymbol{p}_{T}^{miss} + \boldsymbol{p}_{V_{a}} + \boldsymbol{p}_{V_{b}} )\cdot \hat{\boldsymbol{k}}}{\boldsymbol{\beta}_{a} \times\boldsymbol{\beta}_{b} \cdot \hat{\boldsymbol{k}}}\right]  \boldsymbol{\beta}_{b} 
\end{align}

The above relations are independent of the magnitude of the velocity vectors $\boldsymbol{\beta}_{a}$ and $\boldsymbol{\beta}_{b}$ but dependent only on the direction vectors of them,
\begin{align}
\hat{\boldsymbol{r}}_a = \boldsymbol{\beta}_{a}/|\boldsymbol{\beta}_{a}|,\,\,\,\,\,
\hat{\boldsymbol{r}}_b = \boldsymbol{\beta}_{b}/|\boldsymbol{\beta}_{b}|,
\end{align}
thus
\begin{align}
	\boldsymbol{p}_{a} 
	= \left(\frac{\boldsymbol{r}_b \times (\boldsymbol{p}_{T}^{miss} + \boldsymbol{p}_{V_{a}} +  \boldsymbol{p}_{V_{b}}) \cdot \hat{\boldsymbol{k}} }{ \boldsymbol{r}_{b} \times \boldsymbol{r}_{a} \cdot \hat{\boldsymbol{k}}  } \right) \boldsymbol{r}_a, \label{eq:palab}\\
	\boldsymbol{p}_{b} 
	=\left(\frac{\boldsymbol{r}_a \times (\boldsymbol{p}_{T}^{miss} + \boldsymbol{p}_{V_{a}} +  \boldsymbol{p}_{V_{b}}) \cdot \hat{\boldsymbol{k}} }{ \boldsymbol{r}_{a} \times \boldsymbol{r}_{b} \cdot \hat{\boldsymbol{k}}} \right) \boldsymbol{r}_b.
	\label{eq:pblab}
\end{align}
This completes the proof. 


\bibliographystyle{JHEP}
\bibliography{ref}

\end{document}